\documentclass[11pt]{article}
\usepackage{amsmath,amssymb,amsthm,amsxtra,overpic,bbm,bm,epsfig,ulem,color,multirow,appendix}
\usepackage{braket}

\begin{document}

\begin{center}
{\Large A complete study of RGE induced leptogenesis in flavor symmetry scenarios}
\end{center}

\vspace{0.05cm}

\begin{center}
{\bf Zhen-hua Zhao\footnote{zhaozhenhua@lnnu.edu.cn}, Xiang-Yi Wu, Jing Zhang} \\
{ $^1$ Department of Physics, Liaoning Normal University, Dalian 116029, China \\
$^2$ Center for Theoretical and Experimental High Energy Physics, \\ Liaoning Normal University, Dalian 116029, China }
\end{center}

\vspace{0.2cm}

\begin{abstract}
In the literature, motivated by the observed peculiar neutrino mixing pattern and a preliminary experimental hint for maximal Dirac CP phase (i.e., $\delta \sim 3\pi/2$), a lot of flavor and CP symmetries have been proposed to help us understand and explain these experimental results. However, for some flavor-symmetry scenarios (see section~3.1), the leptogenesis mechanism is prohibited to work as usual. To tackle this problem, in this paper we have made an exhausitive study on the possibility that the renormalization group evolution effect may induce a successful leptogenesis for these particular scenarios. Our study provides complementarities to the previous related studies in Refs.~\cite{rgeL1, rgeL2, rgeL3} (see section~3.2).
\end{abstract}

\newpage

\section{Introduction}

As we know, the phenomenon of neutrino oscillations indicates that neutrinos are massive and their flavor eigenstates $\nu^{}_\alpha$ (for $\alpha =e, \mu, \tau$) are certain superpositions of their mass eigenstates $\nu^{}_i$ (for $i =1, 2, 3$) with definite masses $m^{}_i$: $\nu^{}_\alpha = \sum^{}_i U^{}_{\alpha i} \nu^{}_i$ with $U^{}_{\alpha i}$ being the $\alpha i$ element of the $3 \times 3$  neutrino mixing matrix $U$ \cite{xing}. In the standard parametrization, $U$ is expressed in terms of three mixing angles $\theta^{}_{ij}$ (for $ij=12, 13, 23$), one Dirac CP phase $\delta$ and two Majorana CP phases $\rho$ and $\sigma$ as
\begin{eqnarray}
U  = \left( \begin{matrix}
c^{}_{12} c^{}_{13} & s^{}_{12} c^{}_{13} & s^{}_{13} e^{-{\rm i} \delta} \cr
-s^{}_{12} c^{}_{23} - c^{}_{12} s^{}_{23} s^{}_{13} e^{{\rm i} \delta}
& c^{}_{12} c^{}_{23} - s^{}_{12} s^{}_{23} s^{}_{13} e^{{\rm i} \delta}  & s^{}_{23} c^{}_{13} \cr
s^{}_{12} s^{}_{23} - c^{}_{12} c^{}_{23} s^{}_{13} e^{{\rm i} \delta}
& -c^{}_{12} s^{}_{23} - s^{}_{12} c^{}_{23} s^{}_{13} e^{{\rm i} \delta} & c^{}_{23}c^{}_{13}
\end{matrix} \right) \left( \begin{matrix}
e^{{\rm i}\rho} &  & \cr
& e^{{\rm i}\sigma}  & \cr
&  & 1
\end{matrix} \right) \;,
\label{1}
\end{eqnarray}
where the abbreviations $c^{}_{ij} = \cos \theta^{}_{ij}$ and $s^{}_{ij} = \sin \theta^{}_{ij}$ have been employed.

Thanks to the various neutrino oscillation experiments, the three neutrino mixing angles and the neutrino mass squared differences $\Delta m^2_{ij} \equiv m^2_i - m^2_j$ have been measured to a good degree of accuracy, and there is also a preliminary result for $\delta$. Several research groups have performed global analyses of the accumulated neutrino oscillation data to extract the values of these parameters \cite{global,global2}. For definiteness, we will use the results in Ref.~\cite{global} (reproduced in Table~1 here) as reference values in the following numerical calculations. In the light of the large uncertainty of $\delta$, we will treat it as a free parameter.
Note that there are two possible neutrino mass orderings: the normal ordering (NO) case with $m^{}_1 < m^{}_2 < m^{}_3$, and the inverted ordering (IO) case with $m^{}_3 < m^{}_1 < m^{}_2$. On the other hand, neutrino oscillations are completely insensitive to the absolute neutrino mass scale and the Majorana CP phases. Their values can only be inferred from certain non-oscillatory experiments such as the neutrinoless double beta decay experiments \cite{0nbb}. But so far there has not been any lower bound on the lightest neutrino mass, nor any constraint on the Majorana CP phases.

For the neutrino mixing angles, Table~1 shows that $\theta^{}_{12}$ and $\theta^{}_{23}$ are close to some special values (i.e., $\sin^2 \theta^{}_{12} \sim 1/3$ and $\sin^2 \theta^{}_{23} \sim 1/2$). But $\theta^{}_{13}$ is relatively small. In fact, before its value was measured, $\theta^{}_{13}$ had been widely expected to be vanishingly small. For the ideal case of $\sin^2 \theta^{}_{12} = 1/3$, $\sin^2 \theta^{}_{23} = 1/2$ and $\theta^{}_{13} =0$, the neutrino mixing matrix takes a very simple form as
\begin{eqnarray}
U^{}_{\rm TBM}= \displaystyle \frac{1}{\sqrt 6} \left( \begin{array}{ccc}
2 & \sqrt{2} & 0 \cr
1 & - \sqrt{2}  & -\sqrt{3}  \cr
1 & - \sqrt{2}  & \sqrt{3} \cr
\end{array} \right)  \;,
\label{2}
\end{eqnarray}
which is referred to as the tribimaximal (TBM) mixing \cite{TB}. This particular mixing has inspired intensive model-building studies with the employment of certain discrete non-Abelian flavor symmetries (e.g., ${\rm A}^{}_4$ and ${\rm S}^{}_4$) \cite{FS}.
However, the observation of a relatively large $\theta^{}_{13}$ (compared to $0$) motivates us to make corrections for the TBM mixing. In this connection, an economical and natural choice is to retain its first or second column while modifying the other two columns within the unitary constraints, thus yielding the first or second trimaximal (TM1 or TM2) mixing \cite{TM}
\begin{eqnarray}
U^{}_{\rm TM1}=  \displaystyle \frac{1}{\sqrt 6} \left( \begin{array}{ccc}
2 & \cdot & \cdot \cr
1 & \cdot & \cdot \cr
1 & \cdot & \cdot \cr
\end{array} \right)  \;, \hspace{1cm}
U^{}_{\rm TM2}=  \displaystyle \frac{1}{\sqrt 3} \left( \begin{array}{ccc}
\cdot & 1 & \cdot \cr
\cdot & -1 &  \cdot \cr
\cdot &  -1 & \cdot \cr
\end{array} \right)  \;,
\label{3}
\end{eqnarray}
where the dot signs denote the unspecified elements.

In the literature, in order to explain the observed neutrino mixing angles and predict the CP phases at the same time, people have also made a lot of attempts to combine the flavor symmetry with a generalized CP symmetry \cite{gcp}. An attractive and well-studied example of the generalized CP symmetry is the $\mu$-$\tau$ reflection symmetry \cite{MT, MTR}, under which the neutrino mass matrix keeps invariant with respect to the following transformations of three left-handed neutrino fields
\begin{eqnarray}
\nu^{}_{e} \leftrightarrow \nu^{c}_e \;, \hspace{1cm} \nu^{}_{\mu} \leftrightarrow \nu^{c}_{\tau} \;,
\hspace{1cm} \nu^{}_{\tau} \leftrightarrow \nu^{c}_{\mu} \;,
\label{4}
\end{eqnarray}
with the superscript $c$ denoting the charge conjugation of relevant fields.
Then, this symmetry leads to the following interesting predictions for the neutrino mixing angles and CP phases
\begin{eqnarray}
\theta^{}_{23} = \frac{\pi}{4} \;, \hspace{1cm} \delta = \pm \frac{\pi}{2} \;,
\hspace{1cm} \rho = 0 \ {\rm or} \ \frac{\pi}{2} \;, \hspace{1cm} \sigma = 0 \ {\rm or} \ \frac{\pi}{2} \;.
\label{5}
\end{eqnarray}

On the other hand, one of the most popular and natural ways of generating the miniscule neutrino masses is the type-I seesaw model in which at least two heavy right-handed neutrinos $N^{}_I$ ($I=1, 2, 3$) are introduced into the Standard Model (SM) \cite{seesaw}. First of all, $N^{}_I$ can constitute the Yukawa coupling operators together with the left-handed neutrinos $\nu^{}_\alpha$ (which reside in the left-handed lepton doublets $L^{}_\alpha$) and the Higgs doublet $H$: $(Y^{}_{\nu})^{}_{\alpha I} \overline {L^{}_\alpha} H N^{}_I $ with $(Y^{}_{\nu})^{}_{\alpha I}$ being the $\alpha I$ element of the Yukawa coupling matrix $Y^{}_{\nu}$. These operators will generate the Dirac neutrino masses $(M^{}_{\rm D})^{}_{\alpha I}= (Y^{}_{\nu})^{}_{\alpha I} v$ [here $(M^{}_{\rm D})^{}_{\alpha I}$ is the $\alpha I$ element of the Dirac neutrino mass matrix $M^{}_{\rm D}$] after the neutral component of $H$ acquires a nonzero vacuum expectation value (VEV) $v = 174$ GeV. In addition, $N^{}_I$ themselves can also have the Majorana mass terms $\overline{N^c_I} (M^{}_{\rm R})^{}_{IJ} N^{}_J$ [here $(M^{}_{\rm R})^{}_{IJ}$ is the $IJ$ element of the right-handed neutrino mass matrix $M^{}_{\rm R}$].
Then, under the seesaw condition $M^{}_{\rm R} \gg M^{}_{\rm D}$, integrating the right-handed neutrinos out will yield an effective Majorana mass matrix for the three light neutrinos as
\begin{eqnarray}
M^{}_{\nu} = - M^{}_{\rm D} M^{-1}_{\rm R} M^{T}_{\rm D} \;.
\label{6}
\end{eqnarray}
Thanks to such a formula, the smallness of neutrino masses can be naturally explained by the heaviness of right-handed neutrinos. Throughout this paper, without loss of generality, we will work in the basis of $M^{}_{\rm R}$ being diagonal as $D^{}_{\rm R} = {\rm diag}(M^{}_1, M^{}_2, M^{}_3)$ with $M^{}_I$ being the mass of $N^{}_I$ (in the order of $M^{}_1< M^{}_2 < M^{}_3$, without loss of generality).

\begin{table}\centering
  \begin{footnotesize}
    \begin{tabular}{|c|cc|cc|}
     \hline
      & \multicolumn{2}{c|}{Normal Ordering}
      & \multicolumn{2}{c|}{Inverted Ordering }
      \\
      \cline{2-5}
      & bfp $\pm 1\sigma$ & $3\sigma$ range
      & bfp $\pm 1\sigma$ & $3\sigma$ range
      \\
      \cline{1-5}
      \rule{0pt}{4mm}\ignorespaces
       $\sin^2\theta^{}_{12}$
      & $0.303_{-0.011}^{+0.012}$ & $0.270 \to 0.341$
      & $0.303_{-0.011}^{+0.012}$ & $0.270 \to 0.341$
      \\[1mm]
       $\sin^2\theta^{}_{23}$
      & $0.572_{-0.023}^{+0.018}$ & $0.406 \to 0.620$
      & $0.578_{-0.021}^{+0.016}$ & $0.412 \to 0.623$
      \\[1mm]
       $\sin^2\theta^{}_{13}$
      & $0.02203_{-0.00059}^{+0.00056}$ & $0.02029 \to 0.02391$
      & $0.02219_{-0.00057}^{+0.00060}$ & $0.02047 \to 0.02396$
      \\[1mm]
       $\delta/\pi$
      & $1.09_{-0.14}^{+0.23}$ & $0.96 \to 1.33$
      & $1.59_{-0.18}^{+0.15}$ & $1.41\to 1.74$
      \\[3mm]
       $\Delta m^2_{21}/(10^{-5}~{\rm eV}^2)$
      & $7.41_{-0.20}^{+0.21}$ & $6.82 \to 8.03$
      & $7.41_{-0.20}^{+0.21}$ & $6.82 \to 8.03$
      \\[3mm]
       $\Delta m^2_{3\ell}/(10^{-3}~{\rm eV}^2)$
      & $2.511_{-0.027}^{+0.028}$ & $2.428 \to 2.597$
      & $-2.498_{-0.025}^{+0.032}$ & $-2.581 \to -2.408$
      \\[2mm]
      \hline
    \end{tabular}
  \end{footnotesize}
  \caption{The best-fit values, 1$\sigma$ errors and 3$\sigma$ ranges of six neutrino
oscillation parameters extracted from a global analysis of the existing
neutrino oscillation data as of November 2022 \cite{global}, where $\Delta m^2_{3\ell}= \Delta m^2_{31}>0$  for the NO case and $ \Delta m^2_{3\ell}= \Delta m^2_{32}<0$ for the IO case. }
\label{tab1}
\end{table}

Remarkably, as an extra bonus, the seesaw model also provides an attractive explanation (known as the leptogenesis mechanism \cite{leptogenesis, Lreview}) for the baryon-antibaryon asymmetry of the Universe \cite{planck}
\begin{eqnarray}
Y^{}_{\rm B} \equiv \frac{n^{}_{\rm B}-n^{}_{\rm \bar B}}{s} \simeq (8.69 \pm 0.04) \times 10^{-11}  \;,
\label{3}
\end{eqnarray}
where $n^{}_{\rm B}$ ($n^{}_{\rm \bar B}$) denotes the baryon (antibaryon) number density and $s$ is the entropy density. The leptogenesis mechanism works in a way as follows: a lepton-antilepton asymmetry $Y^{}_{\rm L} \equiv (n^{}_{\rm L} - n^{}_{\rm \bar L})/s$ is firstly generated from the out-of-equilibrium and CP-violating decays of the right-handed neutrinos and then partly converted into the baryon-antibaryon asymmetry via the sphaleron processes: $Y^{}_{\rm B} \simeq - cY^{}_{\rm L}$ with $c = 28/79$ or $8/23$ in the SM or MSSM (Minimal Supersymmetric Standard Model) framework \cite{Lreview}.

However, when the above-mentioned flavor symmetries are implemented in the seesaw model, a problem may arise: for some flavor-symmetry scenarios, the leptogenesis mechanism is prohibited to work as usual (see section~3.1 for detailed explanations). To tackle this problem, this paper intends to study the possibility that the renormalization group evolution (RGE) effect may induce a successful leptogenesis for these particular scenarios. The motivation for such a study is twofold:
\begin{itemize}
\item This effect is inevitable and spontaneous, provided that there is a considerable gap between the flavor-symmetry scale and the leptogenesis scale (approximately the right-handed neutrino mass scale).
\item This effect is minimal, in the sense that it does not need to introduce additional flavor-symmetry-breaking parameters.
\end{itemize}

The remaining parts of this paper are organized as follows. In section~2, we will first recapitulate some basic facts about leptogenesis and RGE of the neutrino Yukawa couplings. In section~3, we explain why the leptogenesis mechanism is prohibited to work for some flavor-symmetry scenarios, and clarify the complementarities of our study to previous related studies. In sections~4, 5 and 6, we will perform the study for three distinct flavor-symmetry scenarios that prohibit the leptogenesis mechanism to work before the RGE effects are included (see section~3.1 for their details), respectively. In section~7 we summarize our main results.

\section{Preliminary}

In this section, we first recapitulate some basic facts about leptogenesis and RGE of the neutrino Yukawa couplings.

\subsection{Some basics for leptogenesis relevant for our study}

As is known, depending on the temperature where leptogenesis takes place (approximately the right-handed neutrino masses), there are the following three distinct flavor regimes for leptogenesis \cite{flavor}.
\begin{itemize}
\item Unflavored regime: in the temperature range above $10^{12}$ GeV where the charged-lepton Yukawa $y^{}_\alpha$-related interactions have not yet entered thermal equilibrium, the three lepton flavors are indistinguishable from one another and should be treated in a universal manner.
\item 2-flavor regime: in the temperature range $10^{9}$---$10^{12}$ GeV where the $y^{}_\tau$-related interactions have entered thermal equilibrium, the $\tau$ flavor is distinguishable from the other two flavors which remain indistinguishable from each other so that there are effectively two flavors (i.e., the $\tau$ flavor and a coherent superposition of the $e$ and $\mu$ flavors).
\item 3-flavor regime: in the temperature range below $10^{9}$ GeV where the $y^{}_\mu$-related interactions have also entered thermal equilibrium, all the three lepton flavors are distinguishable from one another so that they should be treated separately.
\end{itemize}

On the other hand, depending on the mass spectrum (to be hierarchical or nearly degenerate) of the right-handed neutrinos, leptogenesis can proceed in the following two distinct ways. In the case that the right-handed neutrino masses are hierarchical, the final baryon asymmetry mainly comes from the lightest right-handed neutrino $N^{}_1$, since its related processes will effectively washout the lepton asymmetry generated from heavier right-handed neutrinos. In the unflavored regime the final baryon asymmetry from $N^{}_I$ is given by \cite{Lreview}
\begin{eqnarray}
Y^{}_{\rm B} = -c r \varepsilon^{}_I \kappa(\widetilde m^{}_I)  \;,
\label{2.1.1}
\end{eqnarray}
where $c = 28/79$ or $8/23$ (in the SM or MSSM framework, respectively) describes the transition efficiency from the lepton asymmetry to the baryon asymmetry via the sphaleron processes, and $r \simeq 4 \times 10^{-3}$ measures the ratio of the equilibrium number density of $N^{}_I$ to the entropy density.
And $\varepsilon^{}_I$ is the total CP asymmetry between the decay rates of $N^{}_I \to L^{}_\alpha + H$ and their CP-conjugate processes $N^{}_I \to \overline{L}^{}_\alpha + \overline{H}$, which is explicitly given by
\begin{eqnarray}
\varepsilon^{}_{I} = \frac{1}{8\pi (M^\dagger_{\rm D}
M^{}_{\rm D})^{}_{II} v^2} \sum^{}_{J \neq I} {\rm Im}\left[
(M^\dagger_{\rm D} M^{}_{\rm D})^{2}_{IJ}\right] {\cal F} \left( \frac{M^2_J}{M^2_I} \right) \; ,
\label{2.1.2}
\end{eqnarray}
with ${\cal F}(x) = \sqrt{x} \{(2-x)/(1-x)+ (1+x) \ln [x/(1+x)] \}$.
It is a sum (over three lepton flavors) of the following flavored CP asymmetries
\begin{eqnarray}
&& \varepsilon^{}_{I \alpha} = \frac{1}{8\pi (M^\dagger_{\rm D}
M^{}_{\rm D})^{}_{II} v^2} \sum^{}_{J \neq I} \left\{ {\rm Im}\left[(M^*_{\rm D})^{}_{\alpha I} (M^{}_{\rm D})^{}_{\alpha J}
(M^\dagger_{\rm D} M^{}_{\rm D})^{}_{IJ}\right] {\cal F} \left( \frac{M^2_J}{M^2_I} \right) \right. \nonumber \\
&& \hspace{1.cm}
+ \left. {\rm Im}\left[(M^*_{\rm D})^{}_{\alpha I} (M^{}_{\rm D})^{}_{\alpha J} (M^\dagger_{\rm D} M^{}_{\rm D})^*_{IJ}\right] {\cal G}  \left( \frac{M^2_J}{M^2_I} \right) \right\} \; ,
\label{2.1.3}
\end{eqnarray}
with ${\cal G}(x) = 1/(1-x)$.
Finally, $\kappa(\widetilde m^{}_I) \leq 1$ is the efficiency factor (i.e., survival probability of the produced lepton asymmetry from the decays of $N^{}_I$) which takes account of the washout effects due to the inverse decays of $N^{}_I$ and various lepton-number-violating scattering processes. Its concrete value is determined by the washout mass parameter
\begin{eqnarray}
\widetilde m^{}_I = \sum^{}_\alpha \widetilde m^{}_{I \alpha} = \sum^{}_\alpha  \frac{|(M^{}_{\rm D})^{}_{\alpha I}|^2}{M^{}_I} \;,
\label{2.1.4}
\end{eqnarray}
and can be numerically calculated by solving relevant Boltzmann equations \cite{Lreview}. In the strong washout regime which applies in most of realistic leptogenesis parameter space, the efficiency factor is roughly inversely proportional to the washout mass parameter.
In the 2-flavor regime, the final baryon asymmetry from $N^{}_I$ is given by
\begin{eqnarray}
Y^{}_{\rm B}
=  -c r \left[ \varepsilon^{}_{I \gamma} \kappa \left( \widetilde m^{}_{I \gamma} \right) + \varepsilon^{}_{I \tau} \kappa \left( \widetilde m^{}_{I \tau} \right) \right]
 \;,
\label{2.1.5}
\end{eqnarray}
with $\varepsilon^{}_{I \gamma} = \varepsilon^{}_{I e} + \varepsilon^{}_{I \mu}$ and $\widetilde m^{}_{I \gamma} = \widetilde m^{}_{I e} + \widetilde m^{}_{I \mu}$. In the 3-flavor regime, the final baryon asymmetry from $N^{}_I$ is given by
\begin{eqnarray}
Y^{}_{\rm B} = -c r \left[ \varepsilon^{}_{I e} \kappa \left( \widetilde m^{}_{I e} \right) + \varepsilon^{}_{I \mu} \kappa \left( \widetilde m^{}_{I \mu} \right) + \varepsilon^{}_{I \tau} \kappa \left(\widetilde m^{}_{I\tau} \right) \right] \; .
\label{2.1.6}
\end{eqnarray}

In the case that the right-handed neutrino masses are nearly degenerate, there will be two important differences for the leptogenesis results. First, the CP asymmetries will get resonantly enhanced as \cite{resonant}
\begin{eqnarray}
\varepsilon^{}_{I\alpha} = \frac{{\rm Im}\left\{ (M^*_{\rm D})^{}_{\alpha I} (M^{}_{\rm D})^{}_{\alpha J}
\left[ M^{}_J (M^\dagger_{\rm D} M^{}_{\rm D})^{}_{IJ} + M^{}_I (M^\dagger_{\rm D} M^{}_{\rm D})^{}_{JI} \right] \right\} }{8\pi  v^2 (M^\dagger_{\rm D} M^{}_{\rm D})^{}_{II}} \cdot \frac{M^{}_I \Delta M^2_{IJ}}{(\Delta M^2_{IJ})^2 + M^2_I \Gamma^2_J} \;,
\label{2.1.7}
\end{eqnarray}
where $\Delta M^2_{IJ} \equiv M^2_I - M^2_J$ has been defined and $\Gamma^{}_J= (M^\dagger_{\rm D} M^{}_{\rm D})^{}_{JJ} M^{}_J/(8\pi v^2)$ is the decay rate of $N^{}_J$ (for $J \neq I$).
In this case, the total CP asymmetry $\varepsilon^{}_{I}$ is obtained as
\begin{eqnarray}
\varepsilon^{}_{I} = \frac{{\rm Im}\left[ (M^\dagger_{\rm D} M^{}_{\rm D})^{2}_{IJ} \right] }{8\pi  v^2 (M^\dagger_{\rm D} M^{}_{\rm D})^{}_{II}} \cdot \frac{M^{}_I M^{}_J \Delta M^2_{IJ}}{(\Delta M^2_{IJ})^2 + M^2_I \Gamma^2_J} \;.
\label{2.1.8}
\end{eqnarray}
Second, the contributions of the nearly degenerate right-handed neutrinos to the final baryon asymmetry will be on the same footing and should be taken into consideration altogether. Correspondingly, the final baryon asymmetry is given by
\begin{eqnarray}
&& {\rm unflavored}: \hspace{0.5cm} Y^{}_{\rm B}  = - c r \kappa \left( \sum \widetilde m^{}_I \right) \sum \varepsilon^{}_{I} \;, \nonumber \\
&& {\rm 2-flavor}: \hspace{0.5cm} Y^{}_{\rm B}  = - c r \left[ \kappa \left( \sum \widetilde m^{}_{I \gamma} \right) \sum \varepsilon^{}_{I \gamma} +
\kappa \left( \sum \widetilde m^{}_{I \tau} \right) \sum \varepsilon^{}_{I \tau}  \right] \;, \nonumber \\
&& {\rm 3-flavor}: \hspace{0.5cm} Y^{}_{\rm B}  = - c r \left[ \kappa \left( \sum \widetilde m^{}_{I e} \right) \sum \varepsilon^{}_{I e} + \kappa \left( \sum \widetilde m^{}_{I \mu} \right) \sum \varepsilon^{}_{I \mu} + \kappa \left( \sum \widetilde m^{}_{I \tau} \right) \sum \varepsilon^{}_{I \tau} \right] \;,
\label{2.1.9}
\end{eqnarray}
where the $\sum$ signs denote the sum over all the nearly degenerate right-handed neutrinos.
Note that, for each lepton flavor, the washout is described by the sum of the related washout mass parameter for each right-handed neutrino.

\subsection{RGE of neutrino Yukawa couplings}

In the literature, given that the origin of the flavor symmetries still remains to be unclear and they may find an origin from some physics near to the Planck scale (e.g., the idea of modular flavor symmetry is inspired by top-down considerations from string theory \cite{modular}), they are usually placed at a very high energy scale $\Lambda^{}_{\rm FS}$ (at least above the seesaw scale in order to be implemented in the seesaw model) \cite{FS}.
When dealing with leptogenesis which takes place around the right-handed neutrino mass scale $M^{}_0$, one should take account of the renormalization group evolution effect if there is a considerable gap between $\Lambda^{}_{\rm FS}$ and $M^{}_0$.
In this paper, we will consider to include the RGE effects within both the SM and MSSM frameworks, given that in the literature many flavor-symmetry models have been formulated in the MSSM framework: in order to break the flavor symmetry in a proper way, one needs to introduce some flavon fields which transform as multiplets of the flavor symmetry and develop particular VEV alignments; and the most
popular approach to derive the desired flavon VEV alignments is
provided by the so-called F-term alignment mechanism which is realized in a supersymmetric setup \cite{FS}. In particular, all of the modular flavor-symmetry models have been formulated in the MSSM framework since the modular flavor symmetry needs supersymmetry to preserve the holomorphicity of the
modular form \cite{modular}.

At the one-loop level, the running behaviour of the Dirac neutrino mass matrix (i.e., the neutrino Yukawa coupling matrix) is described by \cite{ynu}
\begin{eqnarray}
16 \pi^2 \frac{d M^{}_{\rm D}}{dt} = \left[ \frac{3}{2} Y^{}_\nu Y^\dagger_\nu - \frac{3}{2} Y^{}_l Y^\dagger_l + {\rm Tr} \left( 3 Y^{}_u Y^\dagger_u + 3 Y^{}_d Y^\dagger_d +  Y^{}_\nu Y^\dagger_\nu + Y^{}_l Y^\dagger_l \right) - \frac{9}{20} g^2_1 - \frac{9}{4} g^2_2 \right]  M^{}_{\rm D} \;,
\label{2.2.1}
\end{eqnarray}
in the SM framework, and
\begin{eqnarray}
16 \pi^2 \frac{d M^{}_{\rm D}}{dt} = \left[ 3 Y^{}_\nu Y^\dagger_\nu + Y^{}_l Y^\dagger_l + {\rm Tr} \left( 3 Y^{}_u Y^\dagger_u + Y^{}_\nu Y^\dagger_\nu  \right) - \frac{3}{5} g^2_1 - 3 g^2_2 \right]  M^{}_{\rm D} \;,
\label{2.2.2}
\end{eqnarray}
in the MSSM framework. Here $t$ denotes $\ln(\mu/\Lambda^{}_{\rm FS})$ with $\mu$ being the renormalization scale, $Y^{}_{u, d}$ are the up-type-quark and down-type-quark Yukawa matrices and $g^{}_{1, 2}$ are the gauge couplings. In the basis of the charged-lepton mass matrix $M^{}_l$ being diagonal, the Yukawa coupling matrix for three charged leptons
is given by $Y^{}_l = {\rm diag} (y^{}_e, y^{}_\mu, y^{}_\tau)$.

In the SM framework, an integration of Eq.~(\ref{2.2.1}) enables us to obtain the Dirac neutrino mass matrix $M^{}_{\rm D}(M^{}_0)$ at the right-handed neutrino mass scale from its counterpart $M^{}_{\rm D}(\Lambda^{}_{\rm FS})$ at the flavor-symmetry scale as \cite{IRGE}
\begin{eqnarray}
M^{}_{\rm D} (M^{}_0) = I^{}_{0} \left( \begin{array}{ccc}
1+\Delta^{}_{e} &   &  \cr
 & 1 +\Delta^{}_{\mu} &  \cr
 &  &  1+\Delta^{}_{\tau} \cr
\end{array} \right)
M^{}_{\rm D} (\Lambda^{}_{\rm FS}) \;,
\label{2.2.3}
\end{eqnarray}
where
\begin{eqnarray}
&& I^{}_{0}  =  {\rm exp} \left\{ - \frac{1}{16 \pi^2} \int^{\ln (\Lambda^{}_{\rm FS}/M^{}_0)}_{0} \left[ {\rm Tr} \left( 3 Y^{}_u Y^\dagger_u + 3 Y^{}_d Y^\dagger_d +  Y^{}_\nu Y^\dagger_\nu + Y^{}_l Y^\dagger_l \right) - \frac{9}{20} g^2_1 - \frac{9}{4} g^2_2 \right] \ {\rm dt} \right\} \;, \nonumber \\
&& \Delta^{}_{\alpha}   =   \frac{3}{32 \pi^2}\int^{\ln (\Lambda^{}_{\rm FS}/M^{}_0)}_{0} y^2_{\alpha} \ {\rm dt} \simeq \frac{3}{32 \pi^2} y^2_{\alpha} \ln \left(\frac{\Lambda^{}_{\rm FS}}{M^{}_0} \right) \;.
\label{2.2.4}
\end{eqnarray}
One can see that $I^{}_0$ is just an overall rescaling factor and it can be absorbed by the redefinitions of the model parameters and thus can be dropped without affecting the final results. In contrast, in spite of being small, $\Delta^{}_\alpha$ can modify the structure of $M^{}_{\rm D}$, inducing remarkable consequences for leptogenesis as will be seen below. Numerically, one has $\Delta^{}_{e} \ll \Delta^{}_{\mu} \ll \Delta^{}_{\tau} \ll 1$ (as a result of $y^{}_e \ll y^{}_\mu \ll y^{}_\tau \ll 1$), so it is an excellent approximation for us to only keep $\Delta^{}_{\tau}$ in the following calculations. Taking $\Lambda^{}_{\rm FS}/M^{}_0 =100$ as a benchmark value, one has $\Delta^{}_\tau \sim 4.4\times 10^{-6}$. On the other hand, in the MSSM framework, one has $y^{2}_\tau = (1+ \tan^2{\beta}) m^2_\tau/v^2$ and consequently $\Delta^{}_\tau$ can be greatly enhanced by a large $\tan{\beta}$ value. To be explicit, the value of $\Delta^{}_{\tau}$ depends on the value of $\tan{\beta}$ in a way as
\begin{eqnarray}
\Delta^{}_{\tau} \simeq -\frac{1}{16 \pi^2} (1+ \tan^2{\beta}) \frac{m^2_\tau}{v^2} \ln \left(\frac{\Lambda^{}_{\rm FS}}{M^{}_0} \right) \;.
\label{2.2.5}
\end{eqnarray}

\section{Motivations of our study and its complementarities to previous related studies}

In this section, we explain why the leptogenesis mechanism is prohibited to work for some flavor-symmetry scenarios, and clarify the complementarities of our study to previous related studies.

\subsection{Motivations of our study}

In order to make it transparent what kind of flavor-symmetry scenarios will prohibit the leptogenesis mechanism to work, we make use of the Casas-Ibarra parametrization of $M^{}_{\rm D}$ \cite{CI}:
\begin{eqnarray}
M^{}_{\rm D} = {\rm i} U D^{1/2}_\nu O D^{1/2}_{\rm R}  \;,
\label{3.1.1}
\end{eqnarray}
with $D^{1/2}_\nu = {\rm diag}(\sqrt{m^{}_1}, \sqrt{m^{}_2}, \sqrt{m^{}_3})$ and $D^{1/2}_{\rm R} = {\rm diag}(\sqrt{M^{}_1}, \sqrt{M^{}_2}, \sqrt{M^{}_3})$. Here $U$ is just the neutrino mixing matrix while $O$ is a complex orthogonal matrix satisfying $O^T O =I$.

By inserting Eq.~(\ref{3.1.1}) into Eqs.~(\ref{2.1.2}, \ref{2.1.8}), it is direct to verify that $\varepsilon^{}_I$ will be vanishing for the specific class of seesaw models in which the elements of $O$ are either real or purely imaginary \cite{flavored}, corresponding to the following four forms of $O$:
\begin{eqnarray}
O^{}_1 = O^{}_x O^{}_y O^{}_z  \;, \hspace{1cm}
O^{}_2 = O^{}_x O^{\prime}_y O^{\prime}_z \;, \hspace{1cm}
O^{}_3 = O^{\prime}_x O^{}_y O^{\prime}_z \;, \hspace{1cm}
O^{}_4 = O^{\prime}_x O^{\prime}_y O^{}_z \;,
\label{3.1.2}
\end{eqnarray}
with
\begin{eqnarray}
O^{}_x  = \left( \begin{matrix}
1 & 0 & 0 \cr
0 & \cos x  & \sin x \cr
0 & -\sin x & \cos x
\end{matrix} \right) \;, \hspace{1cm}
O^\prime_x  = \left( \begin{matrix}
1 & 0 & 0 \cr
0 & \cosh x  & {\rm i} \sinh x \cr
0 & - {\rm i} \sinh x & \cosh x
\end{matrix} \right) \;, \nonumber \\
O^{}_y = \left( \begin{matrix}
\cos y & 0 & \sin y \cr
0 & 1  & 0 \cr
-\sin y & 0 & \cos y
\end{matrix} \right) \;, \hspace{1cm}
O^\prime_y = \left( \begin{matrix}
\cosh y & 0 & {\rm i}\sinh y \cr
0 & 1  & 0 \cr
-{\rm i} \sinh y & 0 & \cosh y
\end{matrix} \right) \;, \nonumber \\
O^{}_z = \left( \begin{matrix}
\cos z & \sin z & 0 \cr
-\sin z & \cos z  & 0 \cr
0 & 0 & 1
\end{matrix} \right) \;, \hspace{1cm}
O^\prime_z = \left( \begin{matrix}
\cosh z & {\rm i} \sinh z & 0 \cr
-{\rm i} \sinh z & \cosh z  & 0 \cr
0 & 0 & 1
\end{matrix} \right) \;,
\label{3.1.3}
\end{eqnarray}
where $x$, $y$ and $z$ are real parameters. This means that in the unflavored regime the leptogenesis mechanism is prohibited to work for the above four forms of $O$ in both the cases that the right-handed neutrino masses are hierarchical and nearly degenerate. It is interesting to note that this class of seesaw models can be naturally realized in flavor models with residual CP symmetries \cite{rCP} such as the $\mu$-$\tau$ reflection symmetry.

Similarly, it is easy to see from Eqs.~(\ref{2.1.3}, \ref{2.1.7}) that $\varepsilon^{}_{I\alpha}$ will be vanishing for $O=I$ in which case different columns of $M^{}_{\rm D}$ will be orthogonal to one another [i.e., $(M^\dagger_{\rm D} M^{}_{\rm D})^{}_{IJ} =0$]. This means that in all the three flavor regimes the leptogenesis mechanism is prohibited to work for $O=I$ in both the cases that the right-handed neutrino masses are hierarchical and nearly degenerate. It is interesting to note that this class of seesaw models can be naturally realized in the following two classes of flavor-symmetry models: 1) in the flavor-symmetry models based on the sequential dominance \cite{SD}, the columns of $M^{}_{\rm D}$ are proportional to the columns of the neutrino mixing matrix, so they are orthogonal to one another; 2) in the flavor-symmetry models where the Dirac neutrino matrix is proportional to the identity matrix while the right-handed neutrino mass matrix is non-diagonal (for example, some recently popular modular symmetry models just belong to this kind of models \cite{MD}), the former will become proportional to a unitary matrix (whose columns are orthogonal to one another) after one goes back to the basis with the latter being diagonal via a unitary transformation of the right-handed neutrinos. Note that in the former class of models the right-handed neutrino masses are independent parameters, while in the latter class of models the right-handed neutrino masses are inversely proportional to the light neutrino masses.

Finally, in the case that the right-handed neutrino masses are nearly degenerate, it can be seen from Eq.~(\ref{2.1.7}) that $\varepsilon^{}_{I \alpha}$ will be strongly suppressed by $M^{}_I - M^{}_J$ for $O=O^{\prime}_x$, $O^{\prime}_y$ or $O^{\prime}_z$:
\begin{eqnarray}
\varepsilon^{}_{I\alpha} & \propto & {\rm Im}\left\{ (M^*_{\rm D})^{}_{\alpha I} (M^{}_{\rm D})^{}_{\alpha J}
\left[ M^{}_J (M^\dagger_{\rm D} M^{}_{\rm D})^{}_{IJ} + M^{}_I (M^\dagger_{\rm D} M^{}_{\rm D})^{}_{JI} \right] \right\}  \nonumber \\
& = & (M^{}_I - M^{}_J) {\rm Im}\left[ (M^*_{\rm D})^{}_{\alpha I} (M^{}_{\rm D})^{}_{\alpha J}
 (M^\dagger_{\rm D} M^{}_{\rm D})^{}_{JI} \right] \;.
\label{3.1.4}
\end{eqnarray}
This means that in the 2-flavor and 3-flavor regimes the leptogenesis mechanism is strongly suppressed for these three forms of $O$ in the case that the right-handed neutrino masses are nearly degenerate.
Note that this class of seesaw models can be naturally realized in flavor models with a coexistence of residual flavor and CP symmetries \cite{rCP, TMMT}.

\begin{table}\centering
  \begin{footnotesize}
    \begin{tabular}{|c|c|c|c|}
    \hline
       $O$ & possible symmetry origins
      & flavor regimes  & mass spectrum
      \\ \hline
  $O^{}_1$, $O^{}_2$, $O^{}_3$, $O^{}_4$ & CP  & unflavored & hierarchical, nearly degenerate \\ \hline
   $I$ & non-Abelian & unflavored, 2-flavor, 3-flavor & hierarchical, nearly degenerate \\ \hline
  $O^{\prime}_x$, $O^{\prime}_y$, $O^{\prime}_z$ & CP + non-Abelian & 2-flavor, 3-flavor & nearly degenerate \\ \hline
    \end{tabular}
  \end{footnotesize}
  \caption{The forms of $O$ that prohibit the leptogenesis mechanism to work in certain leptogenesis flavor regimes for certain right-handed neutrino mass spectrum, and their possible symmetry origins.   }
\label{tab2}
\end{table}

For the above flavor-symmetry scenarios that prohibit the leptogenesis mechanism to work (see Table~2 for a summary), motivated by the fact that the RGE effect may modify the structure of $M^{}_{\rm D}$ as shown in Eq.~(\ref{2.2.3}) and consequently induce leptogenesis to work, we will study if the RGE induced leptogenesis can successfully reproduce the observed value of $Y^{}_{\rm B}$.

\subsection{Complementarities of our study to previous related studies}

To our knowledge, there have been three papers dedicated to the study on RGE induced leptogenesis for certain flavor-symmetry scenarios \cite{rgeL1, rgeL2, rgeL3}. The complementarities of our study to these papers are clarified as follows:
\begin{itemize}
\item In Ref.~\cite{rgeL1}, for the scenario of $O=O^{}_1$ (whose elements are all real) and the unflavored regime, the authors have studied the RGE induced leptogenesis in the case that the right-handed neutrino masses are hierarchical. There, $O=O^{}_1$ is assumed to hold at low energies and it is the RGE effect between the low energy and the right-handed neutrino mass scale that induces leptogensis to work. In this paper, we will make an analysis complementary to this study from the following three aspects: 1) First, we will assume the special forms of $O$ such as $O=O^{}_1$ to hold at high energies as usually for the flavor-symmetry models in the literature. In this case it is the RGE effect between the high energy and the right-handed neutrino mass scale that induce leptogensis to work. 2) Second, we will also consider the case that the right-handed neutrino masses are nearly degenerate. 3) Third, we will also consider the scenarios of $O=O^{}_2$, $O^{}_3$ and $O^{}_4$ which contain some purely imaginary elements.
\item In Ref.~\cite{rgeL2}, for the scenario of $O=I$ and the 2-flavor and 3-flavor regimes, the authors have studied the RGE induced leptogenesis in the case that the right-handed neutrino masses are nearly degenerate. There, $O=I$ is also assumed to hold at low energies. In this paper, we will make an analysis complementary to this study from the following two aspects: 1) First, we will assume $O=I$ to hold at high energies. 2) Second, we will also consider the case that the right-handed neutrino masses are hierarchical.
\item In Ref.~\cite{rgeL3}, the authors have studied the RGE induced leptogenesis for the scenario of $O=I$ in specific flavor-symmetry models of the TBM and TM mixings. In this paper, we will make an analysis complementary to this study from the following three aspects: 1) First, we will perform a general analysis rather than being restricted into specific flavor-symmetry models. 2) Second, we will also consider the case that the right-handed neutrino masses are nearly degenerate. 3) Third, we will also consider the case that the right-handed neutrino masses are inversely proportional to the light neutrino masses.
\item In addition, we will newly consider the scenarios of $O=O^{\prime}_x$, $O^{\prime}_y$ or $O^{\prime}_z$ in the 2-flavor and 3-flavor regimes in the case that the right-handed neutrino masses are nearly degenerate.
\end{itemize}

\section{Study for scenarios of $O =O^{}_1$, $O^{}_2$, $O^{}_3$ and $O^{}_4$}

As mentioned in section~3.1, for the scenarios of $O =O^{}_1$, $O^{}_2$, $O^{}_3$ and $O^{}_4$, the leptogenesis mechanism is prohibited to work in the unflavored regime (i.e., $\varepsilon^{}_I=0$), in both the cases that the right-handed neutrino masses are hierarchical and nearly degenerate. For these scenarios, in this section we study if the RGE induced leptogenesis can successfully reproduce the observed value of $Y^{}_{\rm B}$. We first perform the study for the case that the right-handed neutrino masses are hierarchical in section~4.1, and then for the case that the right-handed neutrino masses are nearly degenerate in section~4.2. For simplicity and clarity, we will just consider the cases that only one of the parameters $x$, $y$ and $z$ in Eq.~(\ref{3.1.2}) is non-vanishing (i.e., the scenarios of $O=O^{}_x, O^{}_y, O^{}_z, O^{\prime}_x, O^{\prime}_y$ and $O^{\prime}_z$).

\subsection{Study for hierarchical right-handed neutrino masses}

Let us first perform the study for the case that the right-handed neutrino masses are hierarchical.
In the case of $O=O^{}_x$, $N^{}_1$ will decouple from the other two right-handed neutrinos, so it will not contribute to the final baryon asymmetry (but it is responsible for the generation of $m^{}_1$). In this case, the contribution to the final baryon asymmetry mainly comes from the next-to-lightest right-handed neutrino $N^{}_2$---the so-called $N^{}_2$-dominated scenario \cite{N2}. Thanks to the RGE effects as described in section~2.2, a non-zero $\varepsilon^{}_2$ arises as
\begin{eqnarray}
\varepsilon^{}_2 \simeq \Delta^{}_\tau \frac{M^{}_3 (m^{}_2 - m^{}_3) \sqrt{m^{}_2 m^{}_3} \sin 2x}{4\pi v^2 (m^{}_2 \cos^2 x + m^{}_3 \sin^2 x) } {\cal F} \left( \frac{M^2_3}{M^2_2} \right) \Delta^{}_x \; ,
\label{4.1.1}
\end{eqnarray}
with
\begin{eqnarray}
\Delta^{}_x =  c^{}_{23} c^{}_{13} \left[ c^{}_{12} s^{}_{23} \sin \sigma + s^{}_{12} c^{}_{23} s^{}_{13} \sin (\sigma +\delta)  \right] \;.
\label{4.1.2}
\end{eqnarray}
It is natural that the magnitude of $\varepsilon^{}_2$ is directly controlled by $\Delta^{}_\tau$ (given that $\varepsilon^{}_2$ would be completely vanishing without the inclusion of the RGE effects). Before proceeding, it should be noted that the lepton asymmetry generated from $N^{}_2$ decays is subject to the washout effects from the $N^{}_1$-related interactions. In the case that the right-handed neutrino masses are hierarchical, this additional washout factorizes and the final baryon asymmetry is given by \cite{N1wash}
\begin{eqnarray}
Y^{}_{\rm B} = - c r \varepsilon^{}_2 \kappa(\widetilde m^{}_2) e^{- \frac{3 K^{}_1}{8\pi} } \;,
\label{4.1.3}
\end{eqnarray}
where $K^{}_1 \equiv \tilde m^{}_1/m^{}_*$ has been defined (with $m^{}_* \simeq 1.08 \times 10^{-3}$ eV being the so-called equilibrium neutrino mass \cite{Lreview}).

\begin{figure*}
\centering
\includegraphics[width=6.5in]{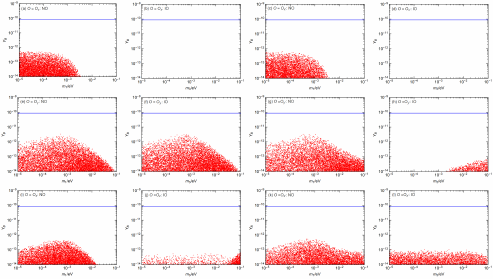}
\caption{ For the scenarios studied in section~4.1, the allowed values of $Y^{}_{\rm B}$ as functions of the lightest neutrino mass $m^{}_1$ and $m^{}_3$ in the NO and IO cases. The blue horizontal line stands for the observed value of $Y^{}_{\rm B}$.}
\label{fig1}
\end{figure*}

For this case, Figure~\ref{fig1}(a) and (b) (for the NO and IO cases, respectively) have shown the allowed values of $Y^{}_{\rm B}$ as functions of the lightest neutrino mass ($m^{}_1$ or $m^{}_3$). These results are obtained for the following parameter settings: for the neutrino mass squared differences and neutrino mixing angles, we employ the data in Table~\ref{tab1}. For $\rho$, $\sigma$, $\delta$ and $x$, we allow them to vary in the range 0---$2\pi$. For $M^{}_2$, we allow it to vary in the range between $10^{12}$ GeV (the lower boundary for the unflavored regime) and $10^{14}$ GeV (above which the $\Delta L=2$ processes mediated by the right-handed neutrinos would greatly suppress the efficiency of leptogenesis \cite{Lreview}). For $M^{}_3$, we allow it to vary in the range $3 M^{}_2$---$10 M^{}_2$ (in fact, the dependence of the final baryon asymmetry on the concrete ratio of $M^{}_3$ and $M^{}_2$ is weak \cite{Lreview}). For $\Lambda^{}_{\rm FS}$, we take it to be $10^{15}$ GeV as a benchmark value (in fact, the dependence of the final results on it is weak since the RGE effects only depend on it in a logarithmic manner). The results show that in the NO case the maximally allowed values of $Y^{}_{\rm B}$ are smaller than the observed value by more than 2 orders of magnitude, and become extremely suppressed for large values of $m^{}_1$. The latter point is simply because one has $\tilde m^{}_1 = m^{}_1$ in the present case so that the washout effects from the $N^{}_1$-related interactions exponentially grow with $m^{}_1$ as shown in Eq.~(\ref{4.1.3}). In the IO case, the allowed values of $Y^{}_{\rm B}$ are suppressed more severely due to that $m^{}_1$ has larger values in this case.

%%%%%%%%%%%%%%%%%%%%%% FIG 1%%%%%%%%%%%%%%%%%%%%%%
%%%%%%%%%%%%%%%%%%%%%%%%%%%%%%%%%%%%%%%%%%%%%%%%%%

In the case of $O=O^{\prime}_x$, one has $\varepsilon^{}_2$ as
\begin{eqnarray}
\varepsilon^{}_2 \simeq - \Delta^{}_\tau \frac{M^{}_3 (m^{}_2 + m^{}_3) \sqrt{m^{}_2 m^{}_3} \sinh 2x}{4\pi v^2 (m^{}_2 \cosh^2 x + m^{}_3 \sinh^2 x) } {\cal F} \left( \frac{M^2_3}{M^2_2} \right) \Delta^{\prime}_x \; ,
\label{4.1.4}
\end{eqnarray}
with
\begin{eqnarray}
\Delta^{\prime}_x =  c^{}_{23} c^{}_{13}  \left[ c^{}_{12} s^{}_{23} \cos \sigma + s^{}_{12} c^{}_{23} s^{}_{13} \cos (\sigma +\delta)  \right] \;.
\label{4.1.5}
\end{eqnarray}
As shown in Figure~\ref{fig1}(c) and (d) (for the NO and IO cases, respectively), the allowed values of $Y^{}_{\rm B}$ in this case are similar to those in the previous case. These results are obtained for the same parameter settings as in the previous case except that here we vary $x$ in the range $-3$---3 for the following consideration: in the cases relevant for $O^\prime_x$, $O^\prime_y$ and $O^\prime_z$, large values of $x$, $y$ and $z$ imply a strong fine tuning because they imply that neutrino masses are much lighter than the individual terms $(M^{}_{\rm D})^{2}_{\alpha I}/M^{}_I$ because of sign cancelations. Therefore, such choices tend to transfer the
explanation of neutrino lightness from the seesaw mechanism to some other mechanism
that has to explain the fine-tuned cancelations. A point of view held by Ref.~\cite{BB} is to consider the $O$ matrices to be "reasonable" if $|\sinh x|, |\sinh y|, |\sinh z| \lesssim 1$ (corresponding to $|x|, |y|, |z| \lesssim 1$), and "acceptable" if $|\sinh x|, |\sinh y|, |\sinh z| \lesssim 10$ (corresponding to $|x|, |y|, |z| \lesssim 3$).

In the case of $O=O^{}_y$, $N^{}_2$ will decouple from the other two right-handed neutrinos, so it will not contribute to the final baryon asymmetry (but it is responsible for the generation of $m^{}_2$). In this case, the contribution to the final baryon asymmetry mainly comes from $N^{}_1$, and the RGE induced non-zero $\varepsilon^{}_1$ arises as
\begin{eqnarray}
\varepsilon^{}_1 \simeq \Delta^{}_\tau \frac{M^{}_3 (m^{}_1 - m^{}_3) \sqrt{m^{}_1 m^{}_3} \sin 2y}{4\pi v^2 (m^{}_1 \cos^2 y + m^{}_3 \sin^2 y) } {\cal F} \left( \frac{M^2_3}{M^2_1} \right) \Delta^{}_y \; ,
\label{4.1.6}
\end{eqnarray}
with
\begin{eqnarray}
\Delta^{}_y =  c^{}_{23} c^{}_{13} \left[ c^{}_{12} c^{}_{23} s^{}_{13} \sin (\delta + \rho)
- s^{}_{12} s^{}_{23}  \sin \rho  \right] \;.
\label{4.1.7}
\end{eqnarray}
In the case of $O=O^{\prime}_y$, one has $\varepsilon^{}_1$ as
\begin{eqnarray}
\varepsilon^{}_1 \simeq - \Delta^{}_\tau \frac{M^{}_3 (m^{}_1 + m^{}_3) \sqrt{m^{}_1 m^{}_3} \sinh 2y}{4\pi v^2 (m^{}_1 \cosh^2 y + m^{}_3 \sinh^2 y) } {\cal F} \left( \frac{M^2_3}{M^2_1} \right) \Delta^{\prime}_y \; ,
\label{4.1.8}
\end{eqnarray}
with
\begin{eqnarray}
\Delta^{\prime}_y =  c^{}_{23} c^{}_{13} \left[ c^{}_{12} c^{}_{23} s^{}_{13} \cos (\delta + \rho)
- s^{}_{12} s^{}_{23}  \cos \rho  \right] \;.
\label{4.1.9}
\end{eqnarray}
For these two cases, the final baryon asymmetry can be calculated according to Eq.~(\ref{2.1.1}) by taking $I=1$. In Figure~\ref{fig1}(e)-(h) (for the cases of $O=O^{}_y$ and $O^\prime_y$ combined with the NO and IO cases, respectively) we have shown the allowed values of $Y^{}_{\rm B}$ as functions of the lightest neutrino mass. The results show that the maximally allowed values of $Y^{}_{\rm B}$ are smaller than the observed value by about 30 times.

In the case of $O=O^{}_z$, $N^{}_3$ will decouple from the other two right-handed neutrinos, so it will not contribute to the final baryon asymmetry (but it is responsible for the generation of $m^{}_3$). In this case, the contribution to the final baryon asymmetry mainly comes from $N^{}_1$, and the RGE induced non-zero $\varepsilon^{}_1$ arises as
\begin{eqnarray}
\varepsilon^{}_1 \simeq \Delta^{}_\tau \frac{M^{}_2 (m^{}_1 - m^{}_2) \sqrt{m^{}_1 m^{}_2} \sin 2z}{4\pi v^2 (m^{}_1 \cos^2 z + m^{}_3 \sin^2 z) } {\cal F} \left( \frac{M^2_2}{M^2_1} \right) \Delta^{}_z \; ,
\label{4.1.10}
\end{eqnarray}
with
\begin{eqnarray}
\Delta^{}_z =  c^{}_{12}  s^{}_{12} (s^2_{23} - c^2_{23} s^2_{13} ) \sin (\rho -\sigma)- c^{}_{23} s^{}_{23} s^{}_{13} \left[ \sin \delta \cos (\rho -\sigma) + (c^2_{12} - s^2_{12}) \cos \delta \sin (\rho -\sigma) \right]  \;.
\label{4.1.11}
\end{eqnarray}
In the case of $O=O^{\prime}_z$, one has $\varepsilon^{}_1$ as
\begin{eqnarray}
\varepsilon^{}_1 \simeq - \Delta^{}_\tau \frac{M^{}_2 (m^{}_1 + m^{}_2) \sqrt{m^{}_1 m^{}_2} \sinh 2z}{4\pi v^2 (m^{}_1 \cosh^2 z + m^{}_2 \sinh^2 z) } {\cal F} \left( \frac{M^2_2}{M^2_1} \right) \Delta^{\prime}_z \; ,
\label{4.1.12}
\end{eqnarray}
with
\begin{eqnarray}
\Delta^{\prime}_z =  c^{}_{12}  s^{}_{12} (s^2_{23} - c^2_{23} s^2_{13} ) \cos (\rho -\sigma)+ c^{}_{23} s^{}_{23} s^{}_{13} \left[ \sin \delta \sin (\rho -\sigma) - (c^2_{12} - s^2_{12}) \cos \delta \cos (\rho -\sigma) \right] \;.
\label{4.1.13}
\end{eqnarray}
For these two cases, Figure~\ref{fig1}(i)-(l) (for the cases of $O=O^{}_z$ and $O^\prime_z$ combined with the NO and IO cases, respectively) have shown the allowed values of $Y^{}_{\rm B}$ as functions of the lightest neutrino mass. The results show that the maximally allowed values of $Y^{}_{\rm B}$ are smaller than the observed value by more than 2 orders of magnitude.

From the above results we see that in the SM framework the allowed values of $Y^{}_{\rm B}$ cannot reach the observed value. On the other hand, in the MSSM framework the final baryon asymmetry can be greatly enhanced for large $\tan \beta$ values: as shown in Eq.~(\ref{2.2.5}), a large $\tan\beta$ value will greatly enhance the size of $\Delta^{}_\tau$ which directly control the strengths of relevant CP asymmetries. In addition, there are the following two issues that should be taken care of in the MSSM framework. 1) The boundaries for different leptogenesis flavor regimes will vary with the value of $\tan \beta$: for example, the boundary for the unflavored and 2-flavor regimes changes from $10^{12}$ GeV to $(1+\tan^2 \beta) 10^{12}$ GeV \cite{MSSM}.
2) In spite of the doubling of the particle spectrum and of the large number of new processes involving superpartners, one does not expect major numerical changes with respect to the non-supersymmetric case. To be specific, for given values of $M^{}_I$, $Y^{}_\nu$ and $Y^{}_l$, the total effect of supersymmetry on the final baryon asymmetry can simply be summarized as a constant factor (for a detailed explanation, see section~10.1 of the third reference in Ref.~\cite{Lreview}):
\begin{eqnarray}
\left. \frac{Y^{\rm MSSM}_{\rm B}}{Y^{\rm SM}_{\rm B}} \right|^{}_{M^{}_I, Y^{}_\nu, Y^{}_l} \simeq \left\{ \begin{array}{l} \sqrt{2} \hspace{0.5cm} ({\rm in \ strong \ washout \ regime} ) \; ; \\ 2 \sqrt{2} \hspace{0.5cm} ({\rm in \ weak \ washout \ regime}) \; . \end{array} \right.
\label{4.1.14}
\end{eqnarray}

%%%%%%%%%%%%%%%%%%%%%% FIG 1%%%%%%%%%%%%%%%%%%%%%%
\begin{figure*}
\centering
\includegraphics[width=6.5in]{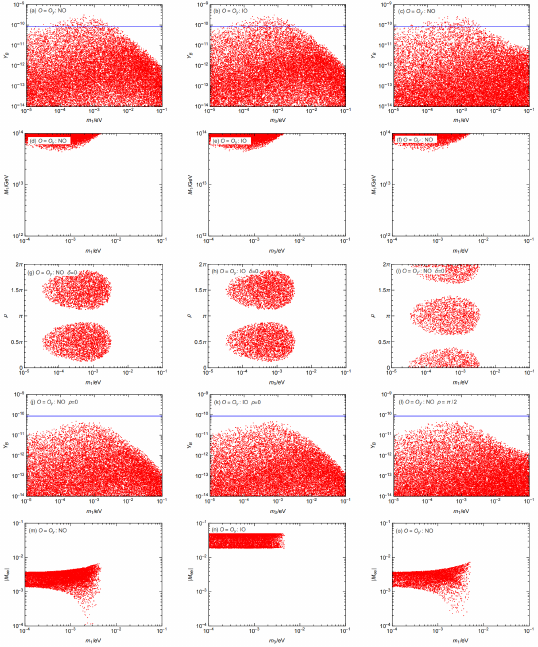}
\caption{ Some further results for the scenarios studied in section~4.1 that allow for a successful leptogenesis in the MSSM framework. (a)-(c): the allowed values of $Y^{}_{\rm B}$ as functions of the lightest neutrino mass;
(d)-(f): the values of $M^{}_1$ (as functions of the lightest neutrino mass) that allow for a successful leptogenesis;
(g)-(i): the relevant parameter space of $\rho$ versus the lightest neutrino mass in the case that $\rho$ is the only source for CP violation; (j)-(l): the allowed values of $Y^{}_{\rm B}$ as functions of the lightest neutrino mass in the case that $\delta$ is the only source for CP violation;  (m)-(o): the allowed values of $|M^{}_{ee}|$ as functions of the lightest neutrino mass in the parameter space for successful leptogenesis. }
\label{fig2}
\end{figure*}
%%%%%%%%%%%%%%%%%%%%%%%%%%%%%%%%%%%%%%%%%%%%%%%%%%

For the scenarios studied in this section, we will only consider $\tan \beta \lesssim 10$ because $\tan \beta \gtrsim 10$ will lift the lower boundary for the unflavored regime to be above $10^{14}$ GeV (above which the $\Delta L=2$ processes mediated by the right-handed neutrinos would greatly suppress the efficiency of leptogenesis). For $\tan \beta \lesssim 10$, the values of $\Delta^{}_\tau$ and consequently $Y^{}_{\rm B}$ can be enhanced by a factor $\lesssim 100$ relative to the corresponding results in the SM framework. In the cases of $O= O^{}_x, O^{\prime}_x, O^{}_z$ and $O^{\prime}_z$, given that in the SM framework the maximally allowed values of $Y^{}_{\rm B}$ are smaller than the observed value by more than 2 orders of magnitude, even in the MSSM framework the observed value of  $Y^{}_{\rm B}$ cannot be reached. On the other hand, in the cases of $O= O^{}_y$ (in both the NO and IO cases) and $O^{\prime}_y$ (in the NO case), the observed value of  $Y^{}_{\rm B}$ can be reached, as shown in Figure~\ref{fig2}(a)-(c) which are obtained by allowing $\tan \beta$ to vary in the range 1---10.
For these cases, in Figure~\ref{fig2}(d)-(f) we have shown the values of $M^{}_1$ (as functions of the lightest neutrino mass) that allow for a successful leptogenesis. The results show that only for $M^{}_1 \gtrsim 5 \times 10^{13}$ GeV can a successful leptogenesis be achieved.
Furthermore, we consider the interesting possibilities that only one of $\delta$, $\rho$ and $\sigma$ is the source for CP violation (e.g., $\delta$ being the only source for CP violation in the case of $\rho=\sigma=0$).
For these possibilities, Figure~\ref{fig2}(g)-(i) have shown the parameter space of $\rho$ versus the lightest neutrino mass for successful leptogenesis in the case that $\rho$ is the only source for CP violation.
One can see that $\rho$ should be around $\pi/2$ or $3\pi/2$ (0 or $\pi$) in the case of $O=O^{}_y$ ($O^\prime_y$), which can be easily understood with the help of Eqs.~(\ref{4.1.7}, \ref{4.1.9}).
But Figure~\ref{fig2}(j)-(l) have shown that the observed value of $Y^{}_{\rm B}$ cannot be reached in the case that $\delta$ is the only source for CP violation (this is because the effects of $\delta$ are always suppressed by the factor $s^{}_{13}$).
Finally, in Figure~\ref{fig2}(m)-(o) we have shown the allowed values of the effective neutrino mass
\begin{eqnarray}
|M^{}_{ee}| = \left| m^{}_1 c^2_{12} c^2_{13} e^{2{\rm i} \rho} + m^{}_2 s^2_{12} c^2_{13}  e^{2{\rm i} \sigma} + m^{}_3 s^2_{13}  e^{-2{\rm i} \delta} \right| \;,
\label{4.1.15}
\end{eqnarray}
that controls the rates of neutrinoless double beta decays \cite{0nbb},
as functions of the lightest neutrino mass in the parameter space for successful leptogenesis. We see that in the NO case it is below 0.006 eV and even might be vanishingly small for $m^{}_1 \sim 0.002$---0.003 eV, which have no chance to be probed by forseeable neutrinoless double beta decay experiments \cite{0nbb}. But in the IO case it is within the range 0.02---0.05 eV, which have the potential to be probed by on-going neutrinoless double beta decay experiments such as LEGEND-200 (with an expected sensitivity for $|M^{}_{ee}|$ in the range 0.027---0.063 eV) \cite{LEGEND}, KamLAND-Zen-800 (with an expected sensitivity for $|M^{}_{ee}|$ in the range 0.038---0.164 eV) \cite{KamL} and SNO+I (with an expected sensitivity for $|M^{}_{ee}|$ in the range 0.031---0.144 eV) \cite{SNO}.

\subsection{Study for nearly degenerate right-handed neutrino masses}

\begin{figure*}
\centering
\includegraphics[width=6.5in]{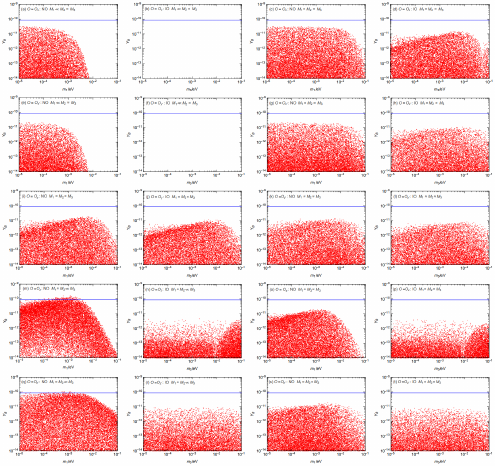}
\caption{ For the scenarios studied in section~4.2, the allowed values of $Y^{}_{\rm B}$ as functions of the lightest neutrino mass. }
\label{fig3}
\end{figure*}

Then, let us perform the study for the case that the right-handed neutrino masses are nearly degenerate.
In the case of $O=O^{}_x$, the contributions to the final baryon asymmetry come from the nearly degenerate $N^{}_2$ and $N^{}_3$, but they will be subject to the washout effects from the $N^{}_1$-related interactions. In this case the RGE induced non-zero $\varepsilon^{}_2$ arises as
\begin{eqnarray}
\varepsilon^{}_2 \simeq \Delta^{}_\tau \frac{(m^{}_3 - m^{}_2)\sqrt {m^{}_2 m^{}_3} \sin 2x}{2\pi v^2 (m^{}_2 \cos^2 x + m^{}_3 \sin^2 x) } \cdot \frac{M^2_0 \Delta M^{}_{32}}{4{{\left( {\Delta M^{}_{32}} \right)}^2} + \Gamma^2_3}\Delta^{}_x\;,
\label{4.2.1}
\end{eqnarray}
where $\Delta^{}_x$ has been given in Eq.~(\ref{4.1.2}), $\Delta M^{}_{IJ} = M^{}_I - M^{}_J$ and $M^{}_0 \approx M^{}_2 \approx M^{}_3$ (like here, in the following we will use $M^{}_0$ to denote the common scale of the two or three nearly degenerate right-handed neutrino masses), while $\varepsilon^{}_3$ can be obtained from it by making the interchange $\cos x \leftrightarrow \sin x$ and the replacement $\Gamma^{}_3 \to \Gamma^{}_2$.
For this case, there are the following two possibilities for the right-handed neutrino mass spectrum: $M^{}_1 \ll M^{}_2 \approx M^{}_3$ and $M^{}_1 \approx M^{}_2 \approx M^{}_3$. For the possibility of $M^{}_1 \ll M^{}_2 \approx M^{}_3$, the final baryon asymmetry is given by
\begin{eqnarray}
Y^{}_{\rm B}  = - c r (\varepsilon^{}_{2} + \varepsilon^{}_{3}) \kappa \left( \widetilde m^{}_2 +  \widetilde m^{}_3 \right)  e^{- \frac{3 K^{}_1}{8\pi} } \;.
\label{4.2.2}
\end{eqnarray}
For this possibility, Figure~\ref{fig3}(a) and (b) (for the NO and IO cases, respectively) have shown the allowed values of $Y^{}_{\rm B}$ as functions of the lightest neutrino mass.
These results are obtained for the same parameter settings as in section~4.1 except that here $M^{}_2$ and $M^{}_3$ are nearly equal to each other and their difference $\Delta M^{}_{32}$ is allowed to vary freely. The results show that in the NO case the maximally allowed values of $Y^{}_{\rm B}$ are smaller than the observed value by about 3 times and in the IO case the allowed values of $Y^{}_{\rm B}$ are severely suppressed.
For the possibility of $M^{}_1 \approx M^{}_2 \approx M^{}_3$, the final baryon asymmetry is given by
\begin{eqnarray}
Y^{}_{\rm B}  = - c r (\varepsilon^{}_{2} + \varepsilon^{}_{3}) \kappa \left( \widetilde m^{}_1 +  \widetilde m^{}_2 +  \widetilde m^{}_3 \right) \;.
\label{4.2.3}
\end{eqnarray}
For this possibility, Figure~\ref{fig3}(c) and (d) (for the NO and IO cases, respectively) have shown the allowed values of $Y^{}_{\rm B}$ as functions of the lightest neutrino mass. The results show that in the NO (IO) case the maximally allowed values of $Y^{}_{\rm B}$ are smaller than the observed value by about 3 times (one order of magnitude).

In the case of $O=O^{\prime}_x$, one has $\varepsilon^{}_2$ as
\begin{eqnarray}
\varepsilon^{}_2 \simeq \Delta^{}_\tau \frac{(m^{}_2 + m^{}_3)\sqrt {m^{}_2 m^{}_3} \sinh 2x}{2\pi v^2 (m^{}_2 \cosh^2 x + m^{}_3 \sinh^2 x) } \cdot \frac{M^2_0 \Delta M^{}_{32}}{4{{\left( {\Delta M^{}_{32}} \right)}^2} + \Gamma^2_3} \Delta^{\prime}_x \;,
\label{4.2.4}
\end{eqnarray}
where $\Delta^{\prime}_x$ has been given in Eq.~(\ref{4.1.5}) and $M^{}_0 \approx M^{}_2 \approx M^{}_3$, while $\varepsilon^{}_3$ can be obtained from it by making the interchange $\cosh x \leftrightarrow \sinh x$ and the replacement $\Gamma^{}_3 \to \Gamma^{}_2$. For this case, Figure~\ref{fig3}(e)-(h) (for the possibilities of $M^{}_1 \ll M^{}_2 \approx M^{}_3$ and $M^{}_1 \approx M^{}_2 \approx M^{}_3$ combined with the NO and IO cases, respectively) have shown the allowed values of $Y^{}_{\rm B}$ as functions of the lightest neutrino mass. The results are quite similar to those for the case of $O=O^{}_x$.

%%%%%%%%%%%%%%%%%%%%%% FIG 1%%%%%%%%%%%%%%%%%%%%%%
%%%%%%%%%%%%%%%%%%%%%%%%%%%%%%%%%%%%%%%%%%%%%%%%%%

In the case of $O=O^{}_y$, the contributions to the final baryon asymmetry come from the nearly degenerate $N^{}_1$ and $N^{}_3$, but they will be subject to the washout effects from the $N^{}_2$-related interactions. In this case the RGE induced non-zero $\varepsilon^{}_1$ arises as
\begin{eqnarray}
\varepsilon^{}_1 \simeq \Delta^{}_\tau \frac{(m^{}_3 - m^{}_1)\sqrt {m^{}_1 m^{}_3} \sin 2y}{2\pi v^2 (m^{}_1 \cos^2 y + m^{}_3 \sin^2 y) } \cdot \frac{M^2_0 \Delta M^{}_{31}}{4{{\left( {\Delta M^{}_{31}} \right)}^2} + \Gamma^2_3}\Delta^{}_y\;,
\label{4.2.5}
\end{eqnarray}
where $\Delta^{}_y$ has been given in Eq.~(\ref{4.1.7}) and $M^{}_0 \approx M^{}_1 \approx M^{}_3$.
In the case of $O=O^{\prime}_y$, one has $\varepsilon^{}_1$ as
\begin{eqnarray}
\varepsilon^{}_1 \simeq \Delta^{}_\tau \frac{(m^{}_1 + m^{}_3)\sqrt {m^{}_1 m^{}_3} \sinh 2y}{2\pi v^2 (m^{}_1 \cosh^2 y + m^{}_3 \sinh^2 y) } \cdot \frac{M^2_0 \Delta M^{}_{31}}{4{{\left( {\Delta M^{}_{31}} \right)}^2} + \Gamma^2_3} \Delta^{\prime}_y \;,
\label{4.2.6}
\end{eqnarray}
where $\Delta^{\prime}_y$ has been given in Eq.~(\ref{4.1.9}) and $M^{}_0 \approx M^{}_1 \approx M^{}_3$.
For these two cases, $\varepsilon^{}_3$ can be obtained from $\varepsilon^{}_1$ by making the interchange $\cos y \leftrightarrow \sin y$ (or $\cosh y \leftrightarrow \sinh y$) and the replacement $\Gamma^{}_3 \to \Gamma^{}_1$.
Note that only the possibility of $M^{}_1 \approx M^{}_2 \approx M^{}_3$ is viable, so the final baryon asymmetry is given by
\begin{eqnarray}
Y^{}_{\rm B}  = - c r (\varepsilon^{}_{1} + \varepsilon^{}_{3}) \kappa \left( \widetilde m^{}_1 +  \widetilde m^{}_2 +  \widetilde m^{}_3 \right) \;.
\label{4.2.7}
\end{eqnarray}
Figure~\ref{fig3}(i)-(l) (for the cases of $O=O^{}_y$ and $O^{\prime}_y$ combined with the NO and IO cases, respectively) have shown the allowed values of $Y^{}_{\rm B}$ as functions of the lightest neutrino mass. The results show that the maximally allowed values of $Y^{}_{\rm B}$ are smaller than the observed value by about one order of magnitude.

\begin{figure*}
\centering
\includegraphics[width=6.5in]{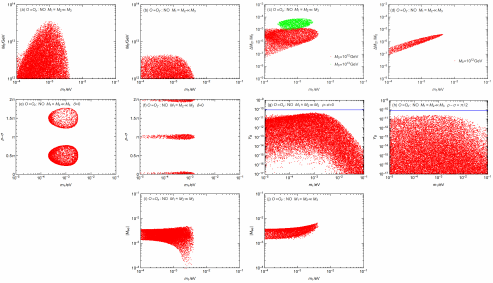}
\caption{ Some further results for the scenarios studied in section~4.2 that allow for a successful leptogenesis.
(a)-(b): the values of $M^{}_0$ (as functions of $m^{}_1$) that allow for a successful leptogenesis; (c)-(d): for the benchmark values of $M^{}_0 = 10^{12}$ (red) and $10^{13}$ (green) GeV, the values of $\Delta M^{}_{21}/M^{}_0$ (as functions of $m^{}_1$) that allow for a successful leptogenesis;
(e)-(f): the relevant parameter space of $\rho-\sigma$ versus $m^{}_1$ in the case that $\rho-\sigma$ is the only source for CP violation; (g)-(h): the allowed values of $Y^{}_{\rm B}$ as functions of $m^{}_1$ in the case that $\delta$ is the only source for CP violation;
(i)-(j): the allowed values of $|M^{}_{ee}|$ as functions of $m^{}_1$ in the parameter space for successful leptogenesis.}
\label{fig4}
\end{figure*}

In the case of $O=O^{}_z$, the contributions to the final baryon asymmetry come from the nearly degenerate $N^{}_1$ and $N^{}_2$, but they may be subject to the washout effects from the $N^{}_3$-related interactions (depending on the mass spectrum of right-handed neutrinos). In this case the RGE induced non-zero $\varepsilon^{}_1$ arises as
\begin{eqnarray}
\varepsilon^{}_1 \simeq \Delta^{}_\tau \frac{(m^{}_2 - m^{}_1)\sqrt {m^{}_1 m^{}_2} \sin 2z}{2\pi v^2 (m^{}_1 \cos^2 z + m^{}_2 \sin^2 z) } \cdot \frac{M^2_0 \Delta M^{}_{21}}{4{{\left( {\Delta M^{}_{21}} \right)}^2} + \Gamma^2_2} \Delta^{}_z\;,
\label{4.2.8}
\end{eqnarray}
where $\Delta^{}_z$ has been given in Eq.~(\ref{4.1.11}) and $M^{}_0 \approx M^{}_1 \approx M^{}_2$.
In the case of $O=O^{\prime}_z$, one has $\varepsilon^{}_1$ as
\begin{eqnarray}
\varepsilon^{}_1 \simeq \Delta^{}_\tau \frac{(m^{}_1 + m^{}_2)\sqrt {m^{}_1 m^{}_2} \sinh 2z}{2\pi v^2 (m^{}_1 \cosh^2 z + m^{}_2 \sinh^2 z) } \cdot \frac{M^2_0 \Delta M^{}_{21}}{4{{\left( {\Delta M^{}_{21}} \right)}^2} + \Gamma^2_2} \Delta^{\prime}_z \;,
\label{4.2.9}
\end{eqnarray}
where $\Delta^{\prime}_z$ has been given in Eq.~(\ref{4.1.13}) and $M^{}_0 \approx M^{}_1 \approx M^{}_2$.
For these two cases, $\varepsilon^{}_2$ can be obtained from $\varepsilon^{}_1$ by making the interchange $\cos z \leftrightarrow \sin z$ (or $\cosh z \leftrightarrow \sinh z$) and the replacement $\Gamma^{}_2 \to \Gamma^{}_1$.
Note that there are the following two possibilities for the right-handed neutrino mass spectrum: $M^{}_1 \approx M^{}_2 \ll M^{}_3$ and $M^{}_1 \approx M^{}_2 \approx M^{}_3$. For the former possibility, the final baryon asymmetry is given by
\begin{eqnarray}
Y^{}_{\rm B}  = - c r (\varepsilon^{}_{1} + \varepsilon^{}_{2}) \kappa \left( \widetilde m^{}_1 +  \widetilde m^{}_2 \right)  \;.
\label{4.2.10}
\end{eqnarray}
For the latter possibility, the final baryon asymmetry is given by
\begin{eqnarray}
Y^{}_{\rm B}  = - c r (\varepsilon^{}_{1} + \varepsilon^{}_{2}) \kappa \left( \widetilde m^{}_1 +  \widetilde m^{}_2 +  \widetilde m^{}_3 \right) \;.
\label{4.2.11}
\end{eqnarray}
Figure~\ref{fig3}(m)-(t) (for the cases of $O=O^{}_z$ and $O^{\prime}_z$ combined with the possibilities of $M^{}_1 \approx M^{}_2 \ll M^{}_3$ and $M^{}_1 \approx M^{}_2 \approx M^{}_3$ and the NO and IO cases, respectively) have shown the allowed values of $Y^{}_{\rm B}$ as functions of the lightest neutrino mass. For the possibility of $M^{}_1 \approx M^{}_2 \ll M^{}_3$, the observed value of $Y^{}_{\rm B}$ can be reached in some parameter region in the NO case (but not in the IO case).
For this possibility, in Figure~\ref{fig4}(a) and (b) we have shown the values of $M^{}_0$ (as functions of $m^{}_1$) that allow for a successful leptogenesis. The results show that in the case of $O=O^{}_z$ ($O^\prime_z$), in order to achieve a successful leptogenesis, $M^{}_0$ should be below $4 \times 10^{13}$ ($4 \times 10^{12}$) GeV.
And in Figure~\ref{fig4}(c) and (d) we have shown the values of $\Delta M^{}_{21}/M^{}_0$ (as functions of $m^{}_1$) that allow for a successful leptogenesis, for the benchmark values of $M^{}_0=10^{12}$ (red) and $10^{13}$ (green) GeV. In the case of $O=O^{}_z$, for $M^{}_0=10^{12}$ GeV, in order to achieve a successful leptogenesis, $\Delta M^{}_{21}/M^{}_0$ should be within the range $10^{-7}$---$10^{-5}$. For $M^{}_0=10^{13}$ GeV, the parameter space for successful leptogenesis shrinks (to be specific, $m^{}_1$ should be within the range 0.0002---0.003 eV and $\Delta M^{}_{21}/M^{}_0$ within the range $10^{-5}$---$10^{-4}$). This is due to that we have fixed $\Lambda^{}_{\rm FS}$ at $10^{15}$ GeV so that the RGE effects (which induce leptogenesis to work) between it and $M^{}_0=10^{13}$ GeV are weaker than those between it and $M^{}_0=10^{12}$ GeV. In the case of $O=O^\prime_z$, for $M^{}_0=10^{12}$ GeV, in order to achieve a successful leptogenesis, $\Delta M^{}_{21}/M^{}_0$ should be within the range $10^{-7}$---$10^{-5}$.
Furthermore, in Figure~\ref{fig4}(e) and (f) we have shown the parameter space of $\rho-\sigma$ versus $m^{}_1$ for successful leptogenesis in the case that $\rho-\sigma$ is the only source for CP violation. One can see that $\rho-\sigma$ should be around $\pi/2$ or $3\pi/2$ (0 or $\pi$) in the case of $O=O^{}_z$ ($O^\prime_z$), which can be easily understood with the help of Eqs.~(\ref{4.1.11}, \ref{4.1.13}). And $m^{}_1$ should be within the range 0.0002---0.003 eV ($\lesssim 0.002$ eV) in the case of $O=O^{}_z$ ($O^\prime_z$).
But as shown in Figure~\ref{fig4}(g) and (h), the observed value of $Y^{}_{\rm B}$ cannot be reached in the case that $\delta$ is the only source for CP violation.
Finally, in Figure~\ref{fig4}(i) and (j) we have shown the allowed values of $|M^{}_{ee}|$ as functions of $m^{}_1$ in the parameter space for successful leptogenesis. We see that in the case of $O=O^{}_z$ it is below 0.006 eV and even might be vanishingly small for $m^{}_1 \sim 0.002$---0.003 eV, and in the case of $O=O^\prime_z$ it is within the range 0.002---0.006 eV, which have no chance to be probed by forseeable neutrinoless double beta decay experiments.
On the other hand, for the possibility of $M^{}_1 \approx M^{}_2 \approx M^{}_3$, the maximally allowed values of $Y^{}_{\rm B}$ are smaller than the observed value by about one order of magnitude. This can be easily understood from the fact that for this possibility the baryon asymmetries generated from $N^{}_1$ and $N^{}_2$ are subject to the additional washout effects from the $N^{}_3$-related interactions.

%%%%%%%%%%%%%%%%%%%%%% FIG 1%%%%%%%%%%%%%%%%%%%%%%
%%%%%%%%%%%%%%%%%%%%%%%%%%%%%%%%%%%%%%%%%%%%%%%%%%

\section{Study for scenarios of $O=O^{\prime}_x$, $O^{\prime}_y$ and $O^{\prime}_z$}

As mentioned in section~3.1, for the scenarios of $O=O^{\prime}_x$, $O^{\prime}_y$ and $O^{\prime}_z$, the CP asymmetries for the right-handed neutrino decays are highly suppressed in the 2-flavor and 3-flavor regimes (i.e., $\varepsilon^{}_{I\alpha} \simeq 0$), in the case that the right-handed neutrino masses are nearly degenerate. For these scenarios, in this section we study if the RGE assisted leptogenesis can successfully reproduce the observed value of $Y^{}_{\rm B}$. We first perform the study for the 2-flavor regime in section~5.1, and then for the 3-flavor regime in section~5.2.

\subsection{Study for 2-flavor regime}

\begin{figure*}
\centering
\includegraphics[width=6.5in]{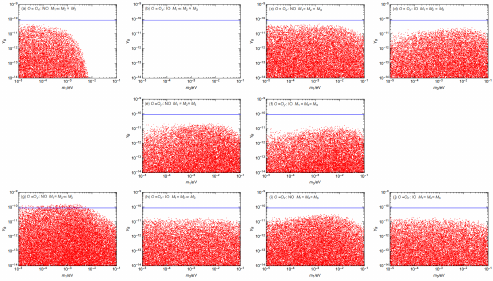}
\caption{ For the scenarios studied in section~5.1, the allowed values of $Y^{}_{\rm B}$ as functions of the lightest neutrino mass. }
\label{fig5}
\end{figure*}

Let us first perform the study for the 2-flavor regime.
In the case of $O=O^{\prime}_x$, the contributions to the final baryon asymmetry come from the nearly degenerate $N^{}_2$ and $N^{}_3$, but they will be subject to the washout effects from the $N^{}_1$-related interactions. In this case the RGE induced contributions to $\varepsilon^{}_{2\alpha}$ are given by
\begin{eqnarray}
&&\hspace{-1.0cm} \varepsilon^{}_{2 e} \simeq \Delta_\tau\frac{ \sqrt {m_2 m_3}}{2\pi v^2(m_2 \cosh^2 x + m_3 \sinh^2 x)} \left[(m_2 c^2_{13} s^2_{12} + m_3 s^2_{13}) \sinh 2x - 2\sqrt {m_2 m_3}  c_{13}  s_{13}  s_{12} \sin (\delta +\sigma) \right. \nonumber \\
&& \hspace{0.1cm} \left. \times \cosh 2x  \right] \cdot \frac{M^2_0\Delta M_{32}}{4{{\left( {\Delta M_{32}} \right)}^2} + \Gamma^2 _3}\Delta^{\prime}_x \;, \nonumber \\
&&\hspace{-1.0cm} \varepsilon^{}_{2 \mu} \simeq \Delta_\tau\frac{  \sqrt {m_2 m_3} }{2\pi v^2(m_2 \cosh^2 x + m_3 \sinh^2 x)  } \{\left[m_2 \left( c^2_{23} c^2_{12} + s^2_{23} s^2_{13} s^2_{12} - 2 c_{23} s_{23} s_{13} c_{12} s_{12} \cos \delta \right) + m_3 s^2_{23} c^2_{13} \right]\nonumber \\
&& \hspace{0.1cm} \times \sinh 2x  +2\sqrt {m_2 m_3}  s_{23}  c_{13} \left[  s_{23} s_{13}  s_{12} \sin (\delta +\sigma)- c_{23}  c_{12} \sin \sigma \right] \cosh 2x  \} \cdot \frac{M^2_0\Delta M_{32}}{4{\left( {\Delta M_{32}} \right)^2} + \Gamma^2 _3}\Delta^{\prime}_x \;, \nonumber \\
&&\hspace{-1.0cm} \varepsilon^{}_{2 \tau} \simeq \Delta_\tau\frac{ \sqrt {m_2 m_3} }{2\pi v^2(m_2 \cosh^2 x + m_3 \sinh^2 x)  } \{ \left[m_2 \left(  s^2_{23} c^2_{12} + c^2_{23} s^2_{13} s^2_{12} + 2 c_{23} s_{23} s_{13} c_{12} s_{12} \cos \delta \right) + m_3 c^2_{23} c^2_{13} \right] \nonumber \\
&& \hspace{0.1cm} \times \sinh 2x + 2\sqrt {m_2 m_3}  c_{23}  c_{13} \left[  c_{23} s_{13}  s_{12} \sin (\delta +\sigma)+ s_{23}  c_{12} \sin \sigma \right] \cosh 2x\} \cdot \frac{M^2_0\Delta M_{32}}{4{\left( {\Delta M_{32}} \right)^2} + \Gamma^2 _3}\Delta^{\prime}_x \;,
\label{5.1.1}
\end{eqnarray}
with $M^{}_0 \approx M^{}_2 \approx M^{}_3$,
while the contributions to $\varepsilon^{}_{3\alpha}$ can be obtained from their $\varepsilon^{}_{2\alpha}$ counterparts by making the interchange $\cosh x \leftrightarrow \sinh x$ and the replacement $\Gamma^{}_3 \to \Gamma^{}_2$.
Then the final baryon asymmetry is given by
\begin{eqnarray}
Y^{}_{\rm B}  = - c r  \left[ (\varepsilon^{}_{2\gamma} + \varepsilon^{}_{3\gamma})  \kappa (\widetilde m^{}_{2 \gamma} + \widetilde m^{}_{3 \gamma} ) e^{- \frac{3 K^{}_{1\gamma}}{8\pi} }  + (\varepsilon^{}_{2\tau} + \varepsilon^{}_{3\tau})  \kappa (\widetilde m^{}_{2 \tau} + \widetilde m^{}_{3 \tau} ) e^{- \frac{3 K^{}_{1\tau}}{8\pi} } \right] \;,
\label{5.1.2}
\end{eqnarray}
for the right-handed neutrino mass spectrum  $M^{}_1 \ll M^{}_2 \approx M^{}_3$, and
\begin{eqnarray}
Y^{}_{\rm B}  = - c r  \left[ (\varepsilon^{}_{2\gamma} + \varepsilon^{}_{3\gamma})  \kappa (\widetilde m^{}_{1 \gamma} + \widetilde m^{}_{2 \gamma} + \widetilde m^{}_{3 \gamma} )  + (\varepsilon^{}_{2\tau} + \varepsilon^{}_{3\tau})  \kappa (\widetilde m^{}_{1 \tau} + \widetilde m^{}_{2 \tau} + \widetilde m^{}_{3 \tau} )  \right] \;,
\label{5.1.3}
\end{eqnarray}
for the right-handed neutrino mass spectrum $M^{}_1 \approx M^{}_2 \approx M^{}_3$, where $K^{}_{1\alpha} \equiv \widetilde m^{}_{1\alpha}/m^{}_*$ has been defined.
Figure~\ref{fig5}(a)-(d) (for the possibilities of $M^{}_1 \ll M^{}_2 \approx M^{}_3$ and $M^{}_1 \approx M^{}_2 \approx M^{}_3$ combined with the NO and IO cases, respectively) have shown the allowed values of $Y^{}_{\rm B}$ as functions of the lightest neutrino mass. These results are obtained for the same parameter settings as in section~4.2 except that here the leptogenesis scale is allowed to vary in the range between $10^9$ GeV and $10^{12}$ GeV where the 2-flavor regime holds. One can see that the results are similar to those in section~4.2. The maximally allowed values of $Y^{}_{\rm B}$ are smaller than the observed value by about 2 times, except that for the possibility of $M^{}_1 \ll M^{}_2 \approx M^{}_3$ the allowed values of $Y^{}_{\rm B}$ are severely suppressed in the IO case.

%%%%%%%%%%%%%%%%%%%%%% FIG 1%%%%%%%%%%%%%%%%%%%%%%
%%%%%%%%%%%%%%%%%%%%%%%%%%%%%%%%%%%%%%%%%%%%%%%%%%

In the case of $O=O^{\prime}_y$, the contributions to the final baryon asymmetry come from the nearly degenerate $N^{}_1$ and $N^{}_3$, but they will be subject to the washout effects from the $N^{}_2$-related interactions. In this case the RGE induced contributions to $\varepsilon^{}_{1\alpha}$ are given by
\begin{eqnarray}
&& \hspace{-1.0cm} \varepsilon^{}_{1 e} \simeq \Delta_\tau \frac{ \sqrt {m_1 m_3} }{2\pi v^2(m_1 \cosh^2 y + m_3 \sinh^2 y)  } \left[(m_1 c^2_{13} c^2_{12} + m_3 s^2_{13}) \sinh 2y  - 2\sqrt {m_1 m_3}  c_{13}  s_{13}  c_{12} \sin (\delta +\rho) \right. \nonumber \\
&& \hspace{0.1cm} \left. \times \cosh 2y \right] \cdot \frac{M^2_0\Delta M_{31}}{4{{\left( {\Delta M_{31}} \right)}^2} + \Gamma^2 _3}\Delta^{\prime}_y \;, \nonumber \\
&&\hspace{-1.0cm} \varepsilon^{}_{1 \mu} \simeq \Delta_\tau\frac{\sqrt {m_1 m_3} }{2\pi v^2(m_1 \cosh^2 y + m_3 \sinh^2 y)  } \{ \left[m_1 \left( c^2_{23} s^2_{12}  + s^2_{23} s^2_{13} c^2_{12} + 2 c_{23} s_{23} s_{13} c_{12} s_{12} \cos \delta \right) + m_3 s^2_{23} c^2_{13} \right]\nonumber \\
&& \hspace{0.1cm} \times \sinh 2y +2\sqrt {m_1 m_3}  s_{23}  c_{13} \left[  s_{23} s_{13}  c_{12} \sin (\delta +\rho)+ c_{23}  s_{12} \sin \rho \right] \cosh 2y \} \cdot \frac{M^2_0\Delta M_{31}}{4{\left( {\Delta M_{31}} \right)^2} + \Gamma^2 _3}\Delta^{\prime}_y \;, \nonumber \\
&&\hspace{-1.0cm} \varepsilon^{}_{1 \tau} \simeq \Delta_\tau\frac{\sqrt {m_1 m_3} }{2\pi v^2(m_1 \cosh^2 y + m_3 \sinh^2 y)  } \{ \left[m_1 \left( s^2_{23} s^2_{12} + c^2_{23} s^2_{13} c^2_{12} - 2 c_{23} s_{23} s_{13} c_{12} s_{12} \cos \delta \right) + m_3 c^2_{23} c^2_{13} \right]\nonumber \\
&& \hspace{0.1cm} \times \sinh 2y +2\sqrt {m_1 m_3}  c_{23}  c_{13} \left[  c_{23} s_{13}  c_{12} \sin (\delta +\rho)- s_{23}  s_{12} \sin \rho \right] \cosh 2y \} \cdot \frac{M^2_0 \Delta M_{31}}{4{\left( {\Delta M_{31}} \right)^2} + \Gamma^2 _3}\Delta^{\prime}_y \;,
\label{5.1.4}
\end{eqnarray}
with $M^{}_0 \approx M^{}_1 \approx M^{}_3$,
while the contributions to $\varepsilon^{}_{3\alpha}$ can be obtained from their $\varepsilon^{}_{1\alpha}$ counterparts by making the interchange $\cosh y \leftrightarrow \sinh y$ and the replacement $\Gamma^{}_3 \to \Gamma^{}_1$.
For this case, only the possibility $M^{}_1 \approx M^{}_2 \approx M^{}_3$ for the right-handed neutrino mass spectrum is viable, and the final baryon asymmetry is given by
\begin{eqnarray}
Y^{}_{\rm B}  = - c r  \left[ (\varepsilon^{}_{1\gamma} + \varepsilon^{}_{3\gamma})  \kappa (\widetilde m^{}_{1 \gamma} + \widetilde m^{}_{2 \gamma} + \widetilde m^{}_{3 \gamma} )  + (\varepsilon^{}_{1\tau} + \varepsilon^{}_{3\tau})  \kappa (\widetilde m^{}_{1 \tau} + \widetilde m^{}_{2 \tau} + \widetilde m^{}_{3 \tau} )  \right] \;.
\label{5.1.5}
\end{eqnarray}
Figure~\ref{fig5}(e) and (f) (for the NO and IO cases, respectively) have shown the allowed values of $Y^{}_{\rm B}$ as functions of the lightest neutrino mass. The results show that the maximally allowed values of $Y^{}_{\rm B}$ are smaller than the observed value by about 4 times.

\begin{figure*}
\centering
\includegraphics[width=6.5in]{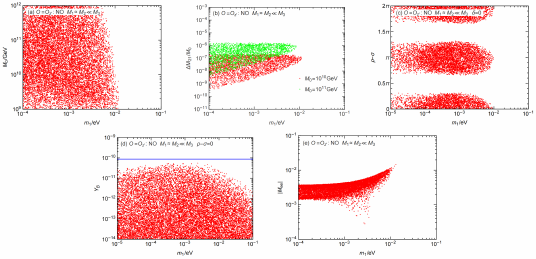}
\caption{ Some further results for the scenarios studied in section~5.1 that allow for a successful leptogenesis.
(a): the values of $M^{}_0$ (as functions of $m^{}_1$) that allow for  a successful leptogenesis; (b) for the benchmark values of $M^{}_0 = 10^{10}$ (red) and $10^{11}$ (green) GeV, the values of $\Delta M^{}_{21}/M^{}_0$ (as functions of $m^{}_1$) that allow for a successful leptogenesis;
(c): the relevant parameter space of $\rho-\sigma$ versus $m^{}_1$ in the case that $\rho-\sigma$ is the only source for CP violation; (d) the allowed values of $Y^{}_{\rm B}$ as functions of $m^{}_1$ in the case that $\delta$ is the only source for CP violation; (e): the allowed values of $|M^{}_{ee}|$ as functions of $m^{}_1$ in the parameter space for successful leptogenesis. }
\label{fig6}
\end{figure*}

In the case of $O=O^{\prime}_z$, the contributions to the final baryon asymmetry come from the nearly degenerate $N^{}_1$ and $N^{}_2$, but they may be subject to the washout effects from the $N^{}_3$-related interactions. In this case the RGE induced contributions to $\varepsilon^{}_{1\alpha}$ are given by
\begin{eqnarray}
&&\hspace{-1.0cm} \varepsilon^{}_{1 e} \simeq \Delta_\tau\frac{ \sqrt {m_1 m_2}  }{2\pi v^2(m_1 \cosh^2 z + m_2 \sinh^2 x)  }  \left[(m_1 c^2_{13} c^2_{12} + m_2 c^2_{13} s^2_{12}) \sinh 2z- 2\sqrt {m_1 m_2}  c^2_{13}  c_{12}  s_{12} \sin (\rho -\sigma) \right. \nonumber \\
&&\hspace{0.1cm} \left. \times \cosh 2z \right] \cdot \frac{M^2_0 \Delta M_{21}}{4{{\left( {\Delta M_{21}} \right)}^2} + \Gamma^2 _2}\Delta^{\prime}_z \;, \nonumber \\
&&\hspace{-1.0cm} \varepsilon^{}_{1 \mu} \simeq \Delta_\tau\frac{\sqrt {m_1 m_2} }{2\pi v^2(m_1 \cosh^2 z + m_2 \sinh^2 z)  } \{ \left[m_1 \left( c^2_{23} s^2_{12} + s^2_{23} s^2_{13} c^2_{12}  + 2 c_{23} s_{23} s_{13} c_{12} s_{12} \cos \delta \right) \right.\nonumber \\
&& \hspace{0.1cm} \left.+ m_2 \left( c^2_{23} c^2_{12} + s^2_{23} s^2_{13} s^2_{12} - 2 c_{23} s_{23} s_{13} c_{12} s_{12} \cos \delta \right)\right]\sinh 2z+ 2\sqrt {m_1 m_2} \left[  c_{12}  s_{12} \left( c^2_{23} - s^2_{13} s^2_{23} \right)  \right.\nonumber \\
&& \hspace{0.1cm} \left. \times \sin (\rho -\sigma)- 2s^2_{12}  c_{23} s_{23} s_{13}  \cos \delta  \sin (\rho -\sigma) +c_{23} s_{23} s_{13} \sin (\delta +\rho -\sigma) \right] \cosh 2z \}\nonumber \\
&& \hspace{0.1cm}\cdot \frac{M^2_0 \Delta M_{21}}{4{\left( {\Delta M_{21}} \right)^2} + \Gamma^2 _2}\Delta^{\prime}_z \;, \nonumber \\
&&\hspace{-1.0cm} \varepsilon^{}_{1 \tau} \simeq \Delta_\tau\frac{ \sqrt {m_1 m_2} }{2\pi v^2(m_1 \cosh^2 z + m_2 \sinh^2 z)  } \{ \left[m_1 \left( s^2_{23} s^2_{12} + c^2_{23} s^2_{13} c^2_{12} - 2 c_{23} s_{23} s_{13} c_{12} s_{12} \cos \delta \right)\right.\nonumber \\
&& \hspace{0.1cm} \left.+ m_2 \left( s^2_{23} c^2_{12} + c^2_{23} s^2_{13} s^2_{12} + 2 c_{23} s_{23} s_{13} c_{12} s_{12} \cos \delta \right)\right]\sinh 2z + 2\sqrt {m_1 m_2} \left[  c_{12}  s_{12} \left( s^2_{23} - c^2_{23} s^2_{13} \right)\right.\nonumber \\
&& \hspace{0.1cm} \left. \times \sin (\rho -\sigma) + 2s^2_{12}  c_{23} s_{23} s_{13}  \cos \delta  \sin (\rho -\sigma)-c_{23} s_{23} s_{13} \sin (\delta +\rho -\sigma) \right] \cosh 2z \}\nonumber \\
&& \hspace{0.1cm}\cdot \frac{M^2_0 \Delta M_{21}}{4{\left( {\Delta M_{21}} \right)^2} + \Gamma^2 _2}\Delta^{\prime}_z \;,
\label{5.1.6}
\end{eqnarray}
with $M^{}_0 \approx M^{}_1 \approx M^{}_2$,
while the contributions to $\varepsilon^{}_{2\alpha}$ can be obtained from their $\varepsilon^{}_{1\alpha}$ counterparts by making the interchange $\cosh z \leftrightarrow \sinh z$ and the replacement $\Gamma^{}_2 \to \Gamma^{}_1$.
Then the final baryon asymmetry is given by
\begin{eqnarray}
Y^{}_{\rm B}  = - c r  \left[ (\varepsilon^{}_{1\gamma} + \varepsilon^{}_{2\gamma})  \kappa (\widetilde m^{}_{1 \gamma} + \widetilde m^{}_{2 \gamma} )  + (\varepsilon^{}_{1\tau} + \varepsilon^{}_{2\tau})  \kappa (\widetilde m^{}_{1 \tau} + \widetilde m^{}_{2 \tau}  )  \right] \;,
\label{5.1.7}
\end{eqnarray}
for the right-handed neutrino mass spectrum $M^{}_1 \approx M^{}_2 \ll M^{}_3$, and
\begin{eqnarray}
Y^{}_{\rm B}  = - c r  \left[ (\varepsilon^{}_{1\gamma} + \varepsilon^{}_{2\gamma})  \kappa (\widetilde m^{}_{1 \gamma} + \widetilde m^{}_{2 \gamma} + \widetilde m^{}_{3 \gamma} )  + (\varepsilon^{}_{1\tau} + \varepsilon^{}_{2\tau})  \kappa (\widetilde m^{}_{1 \tau} + \widetilde m^{}_{2 \tau} + \widetilde m^{}_{3 \tau} )  \right] \;,
\label{5.1.8}
\end{eqnarray}
for the right-handed neutrino mass spectrum $M^{}_1 \approx M^{}_2 \approx M^{}_3$.
Figure~\ref{fig5}(g)-(j) (for the possibilities of $M^{}_1 \approx M^{}_2 \ll M^{}_3$ and $M^{}_1 \approx M^{}_2 \approx M^{}_3$ combined with the NO and IO cases, respectively) have shown the allowed values of $Y^{}_{\rm B}$ as functions of the lightest neutrino mass. For the possibility of $M^{}_1 \approx M^{}_2 \ll M^{}_3$, the observed value of $Y^{}_{\rm B}$ can be reached in some parameter region in the NO case (but not in the IO case).
For this possibility, in Figure~\ref{fig6}(a) we have shown the values of $M^{}_0$ (as functions of $m^{}_1$) that allow for a successful leptogenesis. The results show that a successful leptogenesis can be achieved for $M^{}_0$ in the whole temperature range of the 2-flavor regime (i.e., $10^9$---$10^{12}$ GeV). And in Figure~\ref{fig6}(b) we have shown the values of $\Delta M^{}_{21}/M^{}_0$ (as functions of $m^{}_1$) that allow for a successful leptogenesis, for the benchmark values of $M^{}_0=10^{10}$ (red) and $10^{11}$ (green) GeV. For $M^{}_0=10^{10}$ GeV, in order to achieve a successful leptogenesis, $\Delta M^{}_{21}/M^{}_0$ should be within the range $10^{-9}$---$10^{-7}$. For other values of $M^{}_0$, the values of $\Delta M^{}_{21}/M^{}_0$ that allow for a successful leptogenesis can be obtained from those for $M^{}_0=10^{12}$ GeV by means of a simple rescaling law. For example, the results for $M^{}_0=10^{11}$ GeV (see the figure) can be obtained from those for $M^{}_0=10^{10}$ GeV multiplied by 10 (which is obtained as $10^{11}/10^{10}$). This point can be easily understood from the fact that the results of $Y^{}_{\rm B}$ will keep invariant provided that the combination $\Delta M^{}_{21}/M^2_0$ takes the same value: in the expressions of $\varepsilon^{}_{1\alpha}$ in Eq.~(\ref{5.1.6}), $\Delta M^{}_{21}$ and $M^{}_0$ only take effect in the form of $\Delta M^{}_{21}/M^2_0$ (note that $\Gamma^{}_2$ is proportional to $M^{2}_2 \simeq M^2_0$).
Furthermore, in Figure~\ref{fig6}(c) we have shown the parameter space of $\rho-\sigma$ versus $m^{}_1$ for successful leptogenesis in the case that $\rho-\sigma$ is the only source for CP violation. One can see that $\rho-\sigma$ should be around $0$ or $\pi$, and $m^{}_1$ should be within the range $\lesssim 0.01$ eV.
But as shown in Figure~\ref{fig6}(d), the observed value of $Y^{}_{\rm B}$ cannot be reached in the case that $\delta$ is the only source for CP violation.
Finally, in Figure~\ref{fig6}(e) we have shown the allowed values of $|M^{}_{ee}|$ as functions of $m^{}_1$ in the parameter space for successful leptogenesis. We see that $|M^{}_{ee}|$ is below 0.006 eV and even might be vanishingly small for $m^{}_1 \lesssim 0.004$ eV, which have no chance to be probed by forseeable neutrinoless double beta decay experiments. But it can exceed 0.01 eV for $m^{}_1 \sim 0.01$ eV, which have the potential to be probed by the planned of neutrinoless double beta decay experiments such as LEGEND-1000 (with an expected sensitivity for $|M^{}_{ee}|$ in the range 0.009---0.021 eV) \cite{LEGEND} and nEXO (with an expected sensitivity for $|M^{}_{ee}|$ in the range 0.006---0.027 eV) \cite{EXO}.
On the other hand, for the possibility of $M^{}_1 \approx M^{}_2 \approx M^{}_3$, the maximally allowed values of $Y^{}_{\rm B}$ are smaller than the observed value by about one order of magnitude.

%%%%%%%%%%%%%%%%%%%%%% FIG 1%%%%%%%%%%%%%%%%%%%%%%
%%%%%%%%%%%%%%%%%%%%%%%%%%%%%%%%%%%%%%%%%%%%%%%%%%

\subsection{Study for 3-flavor regime}

\begin{figure*}
\centering
\includegraphics[width=6.5in]{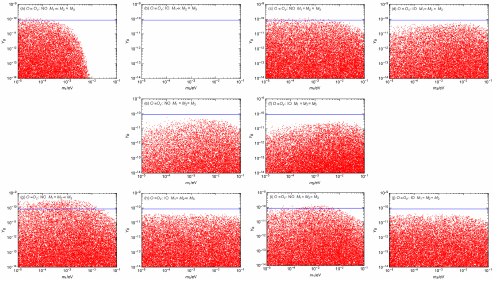}
\caption{ For the scenarios studied in section~5.2, the allowed values of $Y^{}_{\rm B}$ as functions of the lightest neutrino mass. }
\label{fig7}
\end{figure*}

\begin{figure*}
\centering
\includegraphics[width=6.5in]{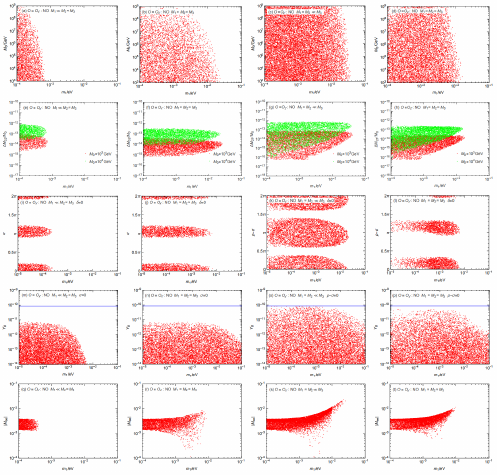}
\caption{ Some further results for the scenarios studied in section~5.2 that allow for a successful leptogenesis.
(a)-(d): the values of $M^{}_0$ (as functions of $m^{}_1$) that allow for a successful leptogenesis; (e)-(h): for the benchmark values of $M^{}_0 = 10^{3}$ (red) and $10^{4}$ (green) GeV, the values of $\Delta M^{}_{32}/M^{}_0$ and $\Delta M^{}_{21}/M^{}_0$ (as functions of $m^{}_1$) that allow for a successful leptogenesis;
(i), (j), (m), (n): the relevant parameter space of $\sigma$ and $\rho-\sigma$ versus $m^{}_1$ in the cases that $\sigma$ and $\rho-\sigma$ are respectively the only source for CP violation; (k), (l), (o), (p): the allowed values of $Y^{}_{\rm B}$ as functions of $m^{}_1$ in the case that $\delta$ is the only source for CP violation;
(q)-(t): the allowed values of $|M^{}_{ee}|$ as functions of $m^{}_1$ in the parameter space for successful leptogenesis. }
\label{fig8}
\end{figure*}

Then, let us perform the study for the 3-flavor regime.
In the case of $O=O^{\prime}_x$, the final baryon asymmetry is given by
\begin{eqnarray}
&& Y^{}_{\rm B}  = - c r \left[ (\varepsilon^{}_{2e} + \varepsilon^{}_{3e})  \kappa (\widetilde m^{}_{2 e} + \widetilde m^{}_{3 e} )  e^{- \frac{3 K^{}_{1e}}{8\pi} } + (\varepsilon^{}_{2\mu} + \varepsilon^{}_{3\mu})  \kappa (\widetilde m^{}_{2 \mu} + \widetilde m^{}_{3 \mu} )  e^{- \frac{3 K^{}_{1\mu}}{8\pi} } \right. \nonumber \\
&& \hspace{1.8cm} \left. + (\varepsilon^{}_{2\tau} + \varepsilon^{}_{3\tau})  \kappa (\widetilde m^{}_{2 \tau} + \widetilde m^{}_{3 \tau} )  e^{- \frac{3 K^{}_{1\tau}}{8\pi} } \right] \; ,
\label{5.2.1}
\end{eqnarray}
for the right-handed neutrino mass spectrum $M^{}_1 \ll M^{}_2 \approx M^{}_3$, and
\begin{eqnarray}
&& Y^{}_{\rm B}  = - c r \left[ (\varepsilon^{}_{2e} + \varepsilon^{}_{3e})  \kappa (\widetilde m^{}_{1 e} + \widetilde m^{}_{2 e} + \widetilde m^{}_{3 e} ) + (\varepsilon^{}_{2\mu} + \varepsilon^{}_{3\mu})  \kappa (\widetilde m^{}_{1 \mu} + \widetilde m^{}_{2 \mu} + \widetilde m^{}_{3 \mu} ) \right. \nonumber \\
&& \hspace{1.8cm} \left. + (\varepsilon^{}_{2\tau} + \varepsilon^{}_{3\tau})  \kappa (\widetilde m^{}_{1 \tau} + \widetilde m^{}_{2 \tau} + \widetilde m^{}_{3 \tau} ) \right] \; ,
\label{5.2.2}
\end{eqnarray}
for the right-handed neutrino mass spectrum $M^{}_1 \approx M^{}_2 \approx M^{}_3$,
where the expressions of $\varepsilon^{}_{I \alpha}$ have been given in last subsection.
Figure~\ref{fig7}(a)-(d) (for the possibilities of $M^{}_1 \ll M^{}_2 \approx M^{}_3$ and $M^{}_1 \approx M^{}_2 \approx M^{}_3$ combined with the NO and IO cases, respectively) have shown the allowed values of $Y^{}_{\rm B}$ as functions of the lightest neutrino mass. These results are obtained for the same parameter settings as in section~5.1 except that here the leptogenesis scale is allowed to vary in the range 1 TeV---$10^9$ GeV where the 3-flavor regime holds, and hence they are quite similar to those in section~5.1 except that the allowed values of $Y^{}_{\rm B}$ get enhanced to some extent. This is simply because in the 3-flavor regime the leptogenesis scale is lower than in the 2-flavor regime so that the RGE effects are relatively stronger (due to the enlargement of the RGE energy gap). For both the possibilities of $M^{}_1 \ll M^{}_2 \approx M^{}_3$ and $M^{}_1 \approx M^{}_2 \approx M^{}_3$, the observed value of $Y^{}_{\rm B}$ can be reached in some parameter region in the NO case (but not in the IO case).
For these possibilities, in Figure~\ref{fig8}(a) and (b) we have shown the values of $M^{}_0$ (as functions of $m^{}_1$) that allow for a successful leptogenesis. The results show that a successful leptogenesis can be achieved for $M^{}_0$ in the whole temperature range of the 3-flavor regime (i.e., $\lesssim 10^{9}$ GeV). And in Figure~\ref{fig8}(e) and (f) we have shown the values of $\Delta M^{}_{32}/M^{}_0$ (as functions of $m^{}_1$) that allow for a successful leptogenesis, for the benchmark values of $M^{}_0=10^{3}$ (red) and $10^{4}$ (green) GeV. For $M^{}_0=10^{3}$ GeV,
in order to achieve a successful leptogenesis, $\Delta M^{}_{32}/M^{}_0$ should be within the range $10^{-15}$---$10^{-13}$. And the results for $M^{}_0=10^{4}$ GeV in relation to those for $M^{}_0=10^{3}$ GeV (i.e., differing by 10 times) do obey the rescaling law mentioned at the end of section~5.1.
Furthermore, in Figure~\ref{fig8}(i) and (j) we have shown the parameter space of $\sigma$ versus $m^{}_1$ for successful leptogenesis in the case that $\sigma$ is the only source for CP violation. One can see that $\sigma$ should be around 0 or $\pi$, and $m^{}_1$ should be within the range $\lesssim 0.0003$ eV ($\lesssim 0.007$ eV) for the possibility of $M^{}_1 \ll M^{}_2 \approx M^{}_3$ ($M^{}_1 \approx M^{}_2 \approx M^{}_3$).
But as shown in Figure~\ref{fig8}(k) and (l), the observed value of $Y^{}_{\rm B}$ cannot be reached in the case that $\delta$ is the only source for CP violation.
Finally, in Figure~\ref{fig8}(q) and (r) we have shown the allowed values of $|M^{}_{ee}|$ as functions of $m^{}_1$ in the parameter space for successful leptogenesis. We see that for the possibility of $M^{}_1 \ll M^{}_2 \approx M^{}_3$ it is within the range 0.001---0.004 eV, which have no chance to be probed by forseeable neutrinoless double beta decay experiments. For the possibility of $M^{}_1 \approx M^{}_2 \approx M^{}_3$ it is below 0.006 eV and even might be vanishingly small for $m^{}_1 \lesssim 0.004$ eV, which have no chance to be probed by forseeable neutrinoless double beta decay experiments. But it can be close to 0.01 eV for $m^{}_1 \sim 0.01$ eV, which have the potential to be probed by the planned of neutrinoless double beta decay experiments such as LEGEND-1000 \cite{LEGEND} and nEXO \cite{EXO}.

%%%%%%%%%%%%%%%%%%%%%% FIG 1%%%%%%%%%%%%%%%%%%%%%%
%%%%%%%%%%%%%%%%%%%%%%%%%%%%%%%%%%%%%%%%%%%%%%%%%%

%%%%%%%%%%%%%%%%%%%%%% FIG 1%%%%%%%%%%%%%%%%%%%%%%
%%%%%%%%%%%%%%%%%%%%%%%%%%%%%%%%%%%%%%%%%%%%%%%%%%

In the case of $O=O^{\prime}_y$, for the only viable right-handed neutrino mass spectrum $M^{}_1 \approx M^{}_2 \approx M^{}_3$, the final baryon asymmetry is given by
\begin{eqnarray}
&& Y^{}_{\rm B}  = - c r \left[ (\varepsilon^{}_{1e} + \varepsilon^{}_{3e})  \kappa (\widetilde m^{}_{1 e} + \widetilde m^{}_{2 e} + \widetilde m^{}_{3 e} ) + (\varepsilon^{}_{1\mu} + \varepsilon^{}_{3\mu})  \kappa (\widetilde m^{}_{1 \mu} + \widetilde m^{}_{2 \mu} + \widetilde m^{}_{3 \mu} ) \right. \nonumber \\
&& \hspace{1.8cm} \left. + (\varepsilon^{}_{1\tau} + \varepsilon^{}_{3\tau})  \kappa (\widetilde m^{}_{1 \tau} + \widetilde m^{}_{2 \tau} + \widetilde m^{}_{3 \tau} ) \right] \; .
\label{5.2.3}
\end{eqnarray}
Figure~\ref{fig7}(e) and (f) (for the NO and IO cases, respectively) have shown the allowed values of $Y^{}_{\rm B}$ as functions of the lightest neutrino mass. The results show that the maximally allowed values of $Y^{}_{\rm B}$ are close to but fail to reach the observed value.

In the case of $O=O^{\prime}_z$, the final baryon asymmetry is given by
\begin{eqnarray}
&& Y^{}_{\rm B}  = - c r \left[ (\varepsilon^{}_{1e} + \varepsilon^{}_{2e})  \kappa (\widetilde m^{}_{1 e} + \widetilde m^{}_{2 e} ) + (\varepsilon^{}_{1\mu} + \varepsilon^{}_{2\mu})  \kappa (\widetilde m^{}_{1 \mu} + \widetilde m^{}_{2 \mu} ) \right. \nonumber \\
&& \hspace{1.8cm} \left. + (\varepsilon^{}_{1\tau} + \varepsilon^{}_{2\tau})  \kappa (\widetilde m^{}_{1 \tau} + \widetilde m^{}_{2 \tau}  ) \right] \; ,
\label{5.2.4}
\end{eqnarray}
for the right-handed neutrino mass spectrum $M^{}_1 \approx M^{}_2 \ll M^{}_3$, and
\begin{eqnarray}
&& Y^{}_{\rm B}  = - c r \left[ (\varepsilon^{}_{1e} + \varepsilon^{}_{2e})  \kappa (\widetilde m^{}_{1 e} + \widetilde m^{}_{2 e} + \widetilde m^{}_{3 e} ) + (\varepsilon^{}_{1\mu} + \varepsilon^{}_{2\mu})  \kappa (\widetilde m^{}_{1 \mu} + \widetilde m^{}_{2 \mu} + \widetilde m^{}_{3 \mu} ) \right. \nonumber \\
&& \hspace{1.8cm} \left. + (\varepsilon^{}_{1\tau} + \varepsilon^{}_{2\tau})  \kappa (\widetilde m^{}_{1 \tau} + \widetilde m^{}_{2 \tau} + \widetilde m^{}_{3 \tau} ) \right] \; ,
\label{5.2.5}
\end{eqnarray}
for the right-handed neutrino mass spectrum $M^{}_1 \approx M^{}_2 \approx M^{}_3$.
Figure~\ref{fig7}(g)-(j) (for the possibilities of $M^{}_1 \approx M^{}_2 \ll M^{}_3$ and $M^{}_1 \approx M^{}_2 \approx M^{}_3$ combined with the NO and IO cases, respectively) have shown the allowed values of $Y^{}_{\rm B}$ as functions of the lightest neutrino mass. For both the possibilities of $M^{}_1 \approx M^{}_2 \ll M^{}_3$ and $M^{}_1 \approx M^{}_2 \approx M^{}_3$, the observed value of $Y^{}_{\rm B}$ can be reached in some parameter region in the NO case (but not in the IO case).
For these possibilities, in Figure~\ref{fig8}(c) and (d) we have shown the values of $M^{}_0$ (as functions of $m^{}_1$) that allow for a successful leptogenesis. The results show that a successful leptogenesis can be achieved for $M^{}_0$ in the whole temperature range of the 3-flavor regime (i.e., $\lesssim 10^{9}$ GeV). And in Figure~\ref{fig8}(g) and (h) we have shown the values of $\Delta M^{}_{21}/M^{}_0$ (as functions of $m^{}_1$) that allow for a successful leptogenesis, for the benchmark values of $M^{}_0=10^{3}$ (red) and $10^{4}$ (green) GeV. For $M^{}_0=10^{3}$ ($10^{4}$) GeV,
in order to achieve a successful leptogenesis, $\Delta M^{}_{21}/M^{}_0$ should be within the range $10^{-16}$---$10^{-13}$ ($10^{-15}$---$10^{-12}$).
Furthermore, in Figure~\ref{fig8}(m) and (n) we have shown the parameter space of $\rho-\sigma$ versus $m^{}_1$ for successful leptogenesis in the case that $\rho-\sigma$ is the only source for CP violation. One can see that $\rho-\sigma$ should be around $0$ or $\pi$, and $m^{}_1$ should be within the range $\lesssim 0.02$ eV ($\lesssim 0.006$ eV) for the possibility of $M^{}_1 \approx M^{}_2 \ll M^{}_3$ ($M^{}_1 \approx M^{}_2 \approx M^{}_3$). But as shown in Figure~\ref{fig8}(o) and (p), the maximally allowed values of $Y^{}_{\rm B}$ cannot reach (despite being very close to) the observed value in the case that $\delta$ is the only source for CP violation.
Finally, in Figure~\ref{fig8}(s) and (t) we have shown the allowed values of $|M^{}_{ee}|$ as functions of $m^{}_1$ in the parameter space for successful leptogenesis. We see that it is below 0.006 eV and even might be vanishingly small for $m^{}_1 \lesssim 0.004$ eV, which have no chance to be probed by forseeable neutrinoless double beta decay experiments. But it can exceed 0.01 eV for $m^{}_1 \sim 0.01$ eV, which have the potential to be probed by the planned of neutrinoless double beta decay experiments such as LEGEND-1000 \cite{LEGEND} and nEXO \cite{EXO}.

\section{Study for scenario of $O =I$}

As mentioned in section~3.1, for the scenario of $O=I$, the leptogenesis mechanism is prohibited to work in all the three flavor regimes, in both the cases that the right-handed neutrino masses are hierarchical and nearly degenerate. For this scenario, in this section we study if the RGE induced leptogenesis can successfully reproduce the observed value of $Y^{}_{\rm B}$. We first perform the study for the case that the right-handed neutrino masses are hierarchical in section~6.1, then for the case that the right-handed neutrino masses are nearly degenerate in section~6.2, and finally for the case that the right-handed neutrino masses are inversely proportional to the light neutrino masses in section~6.3.

\subsection{Study for hierarchical right-handed neutrino masses}

In the case that the right-handed neutrino masses are hierarchical, the final baryon asymmetry mainly comes from $N^{}_1$. In the unflavored regime, the RGE induced non-zero $\varepsilon^{}_1$ arises as
\begin{eqnarray}
\varepsilon^{}_{1} \simeq \Delta^2_\tau \frac{ 1 }{\pi v^2} \left[ M_2 m_2 \Delta_y\Delta^{\prime}_y {\cal F} \left(  \frac{M^2_2}{M^2_1} \right) + M_3 m_3 \Delta_z\Delta^{\prime}_z {\cal F} \left(  \frac{M^2_3}{M^2_1} \right)\right] \;.
\label{6.1.1}
\end{eqnarray}
In this case, since $\varepsilon^{}_{1}$ is suppressed by $\Delta^2_\tau$, the observed value of $Y^{}_{\rm B}$ cannot be reached.

\begin{figure*}
\centering
\includegraphics[width=6.5in]{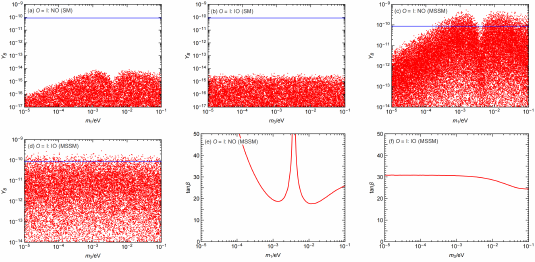}
\caption{ Some results for the scenario studied in section~6.1. (a)-(b): the allowed values of $Y^{}_{\rm B}$ as functions of the lightest neutrino mass in the SM framework; (c)-(d): the allowed values of $Y^{}_{\rm B}$ as functions of the lightest neutrino mass in the MSSM framework; (e)-(f): the minimal values of $\tan \beta$ needed to accommodate a successful leptogenesis.  }
\label{fig9}
\end{figure*}

In the flavored regimes, the RGE induced non-zero $\varepsilon^{}_{1\alpha}$ arise as
\begin{eqnarray}
&&\hspace{-1.0cm} \varepsilon^{}_{1 e} \simeq \Delta_\tau \frac{ c^2_{13} c_{12}}{4\pi v^2} \left\{ M_2 m_2 s_{12}\left[  c_{12}  s_{12} \left( s^2_{23} - c^2_{23} s^2_{13} \right) \sin 2(\rho -\sigma) + c_{23} s_{23} s_{13} \left(s^2_{12}-c^2_{12}\right) \cos \delta  \sin 2(\rho -\sigma)\right. \right. \nonumber \\
&&\hspace{0.1cm}\left.  -c_{23} s_{23} s_{13} \sin \delta  \cos 2(\rho -\sigma) \right]{\cal F} \left(  \frac{M^2_2}{M^2_1} \right)+ M_3 m_3c_{23} s_{13}\left[ c_{23} s_{13} c_{12}\sin 2(\delta + \rho) - s_{23} s_{12}\sin (\delta + 2\rho) \right] \nonumber \\
&&\hspace{0.1cm} \left. \times {\cal F} \left(  \frac{M^2_3}{M^2_1} \right)+ c_{23}  s_{23}  s_{12} s_{13} \sin \delta \left[ M_2 m_2  {\cal G} \left(  \frac{M^2_2}{M^2_1} \right)- M_3 m_3 {\cal G} \left(  \frac{M^2_3}{M^2_1} \right) \right] \right\}\;, \nonumber \\
&&\hspace{-1.0cm} \varepsilon^{}_{1 \mu} \simeq \Delta_\tau \frac{1}{4\pi v^2} \left\{ M_2 m_2 c^{}_{23} s^{}_{23} s^{}_{13} \left[  c^4_{12}  c^{}_{23}  s^{}_{23} s^{}_{13} \sin 2(\delta +\rho - \sigma) -  s^4_{12}  c^{}_{23}  s^{}_{23} s^{}_{13} \sin 2(\delta -\rho + \sigma) + c^3_{12}  s_{12}\right. \right. \nonumber \\
&&\hspace{0.1cm}\left. \times \left(c^2_{23} - s^2_{23}  \right)\left(1 + s^2_{13}  \right) \sin (\delta +2\rho - 2\sigma)  + s^3_{12}  c_{12}\left(c^2_{23} - s^2_{23}  \right)\left(1 + s^2_{13}  \right) \sin (\delta -2\rho + 2\sigma) \right. \nonumber \\
&&\hspace{0.1cm} \left. + c^2_{12} s^2_{12}\left( s^2_{13}- c^2_{23} s^2_{23} - s^4_{13} c^2_{23}s^2_{23}-4c^2_{23}s^2_{23}s^2_{13}  \right)\sin 2(\rho - \sigma) \right]  {\cal F} \left(  \frac{M^2_2}{M^2_1} \right) + M_3 m_3 c^2_{13}c_{23} s_{23}   \nonumber \\
&&\hspace{0.1cm} \times \left[s^2_{12}c_{23} s_{23}\sin 2\rho -s^2_{13}c^2_{12}c_{23} s_{23}\sin 2(\delta +\rho )- s_{13}c_{12} s_{12}\left( c^2_{23} - s^2_{23}\right)\sin (\delta +2\rho )\right]  {\cal F} \left(  \frac{M^2_3}{M^2_1} \right) \nonumber \\
&&\hspace{0.1cm} \left. - c^2_{13}c_{23} s_{23} s_{13}  c_{12}s_{12}\sin \delta \left[ M_2 m_2  {\cal G} \left(  \frac{M^2_2}{M^2_1} \right) - M_3 m_3 {\cal G} \left(  \frac{M^2_3}{M^2_1} \right) \right] \right\}\;, \nonumber \\
&&\hspace{-1.0cm} \varepsilon^{}_{1 \tau} \simeq \Delta_\tau \frac{1}{4\pi v^2} \{ M_2 m_2 c^{}_{23} s^{}_{23} s^{}_{13} \left[  s^4_{12}  c^{}_{23} s^{}_{23} s^{}_{13} \sin 2(\delta -\rho + \sigma) -  c^4_{12}  c^{}_{23} s^{}_{23} s^{}_{13} \sin 2(\delta +\rho - \sigma) - 2 c^3_{12}   s_{12}\right. \nonumber \\
&&\hspace{0.1cm}\left.\left(c^2_{23} s^2_{13} - s^2_{23}  \right)\sin (\delta +2\rho - 2\sigma) - 2 s^3_{12} c_{12} \left(c^2_{23} s^2_{13} - s^2_{23}  \right) \sin (\delta -2\rho + 2\sigma) - c^2_{12} s^2_{12}\left(  c^4_{23}  s^4_{13}+ s^4_{23}\right. \right.\nonumber \\
&&\hspace{0.1cm}\left.\left. - 4 c^2_{23}s^2_{23} s^2_{13}  \right)\sin 2(\rho - \sigma) \right] {\cal F} \left(  \frac{M^2_2}{M^2_1} \right) + M_3 m_3 c^2_{23} c^2_{13} \left[2 c_{23} s_{23} s_{13} c_{12} s_{12} \sin (\delta +2\rho )- s^2_{23} s^2_{12}\sin 2\rho \right. \nonumber \\
&&\hspace{0.1cm} \left. \left.-  c^2_{23} s^2_{13}  c^2_{12} \sin 2(\delta + \rho)\right]  {\cal F} \left(  \frac{M^2_3}{M^2_1} \right) \right \} \;.
\label{6.1.2}
\end{eqnarray}
Then the final baryon asymmetry can be calculated according to Eq.~(\ref{2.1.5}) or Eq.~(\ref{2.1.6}) by taking $I=1$ in the 2-flavor or 3-flavor regime.
For the 2-flavor regime, Figure~\ref{fig9}(a) and (b) (for the NO and IO cases, respectively) have shown the allowed values of $Y^{}_{\rm B}$ as functions of the lightest neutrino mass. The results show that the maximally allowed values of $Y^{}_{\rm B}$ are smaller than the observed value by about 5 orders of magnitude. For the 3-flavor regime, the maximally allowed values of $Y^{}_{\rm B}$ are much smaller (due to the lowering of the leptogenesis scale), so we shall not present them.

%%%%%%%%%%%%%%%%%%%%%% FIG 1%%%%%%%%%%%%%%%%%%%%%%
%%%%%%%%%%%%%%%%%%%%%%%%%%%%%%%%%%%%%%%%%%%%%%%%%%

\begin{figure*}
\centering
\includegraphics[width=6.5in]{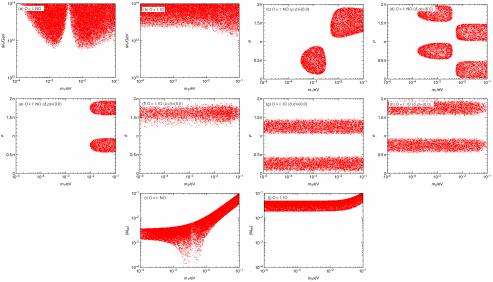}
\caption{ Some further results for the scenario studied in section~6.1 that allows for a successful leptogenesis in the MSSM framework.
(a)-(b): the values of $M^{}_1$ (as functions of the lightest neutrino mass) that allow for a successful leptogenesis.
(c)-(h): the relevant parameter space of $\delta$, $\rho$ and $\sigma$ versus the lightest neutrino mass in the cases that $\delta$, $\rho$ and $\sigma$ are respectively the only source for CP violation; (i)-(j): the allowed values of $|M^{}_{ee}|$ as functions of the lightest neutrino mass in the parameter space for successful leptogenesis.}
\label{fig10}
\end{figure*}

In contrast, in the MSSM framework the observed value of $Y^{}_{\rm B}$ may be successfully reproduced due to the following two enhancement effects from a large $\tan \beta$ value: on the one hand, as shown in Eq.~(\ref{2.2.5}), a large $\tan\beta$ value will greatly enhance the size of $\Delta^{}_\tau$ which directly control the strengths of relevant CP asymmetries; on the other hand, a large $\tan\beta$ value will lift the upper boundary for the two-flavor leptogenesis regime from $10^{12}$ GeV to $(1+\tan^2 \beta) 10^{12}$ GeV, via which the enlargement of the allowed right-handed neutrino mass scale will also enhance the sizes of relevant CP asymmetries. Figure~\ref{fig9}(c) and (d) (for the NO and IO cases, respectively) have shown the allowed values of $Y^{}_{\rm B}$ as functions of the lightest neutrino mass. These results are obtained by allowing $\tan \beta$ to vary in the range 1---50 and correspondingly $M^{}_1$ to vary in the range between $(1+\tan^2 \beta)10^{9}$ GeV and $(1+\tan^2 \beta)10^{12}$ GeV (in order for the 2-flavor regime to hold). Note that $M^{}_1$ is always restricted to be $\lesssim 10^{14}$ GeV (in order to avoid strong washout effects due to the $\Delta L=2$ processes). The results show that the observed value of $Y^{}_{\rm B}$ can be reached in some parameter region. And Figure~\ref{fig9}(e) and (f) (for the NO and IO cases, respectively) have shown the minimal values of $\tan \beta$ needed to accommodate a successful leptogenesis. One can see that $\tan \beta$ should be $\gtrsim 20$ ($\gtrsim 30$) in the NO (IO) case in order to accommodate a successful leptogenesis.
In addition, in Figure~\ref{fig10}(a) and (b) we have shown the values of $M^{}_1$ (as functions of the lightest neutrino mass) that allow for a successful leptogenesis. The results show that a successful leptogenesis can be achieved for $M^{}_1 \gtrsim 5 \times 10^{12}$ ($10^{13}$) GeV in the NO (IO) case.
Furthermore, in Figure~\ref{fig10}(c)-(h) we have shown the parameter space of $\delta$, $\rho$ and $\sigma$ versus the lightest neutrino mass for successful leptogenesis in the case that $\delta$, $\rho$ and $\sigma$ are respectively the only source for CP violation. In the case that $\delta$ is the only source for CP violation, in order to accommodate a successful leptogenesis, in the NO case $\delta$ should be around $\pi/2$ or $3\pi/2$ and $m^{}_1$ should be $\gtrsim 0.0003$ eV, while in the IO case $\delta$ should be around $3\pi/2$ and $m^{}_3$ can take arbitrary values. In the case that $\rho$ is the only source for CP violation, in order to accommodate a successful leptogenesis, in the NO case $\rho$ should be around $\pi/4$, $3\pi/4$, $5\pi/4$ or $7\pi/4$ and $m^{}_1$ should be $\gtrsim 0.0002$ eV, while in the IO case $\rho$ should be around $\pi/4$ or $3\pi/4$. In the case that $\sigma$ is the only source for CP violation, in order to accommodate a successful leptogenesis, in the NO case $\sigma$ should be around $3\pi/4$ or $7\pi/4$ and $m^{}_1$ should be $\gtrsim 0.01$ eV, while in the IO case $\sigma$ should also be around $3\pi/4$ or $7\pi/4$.
Finally, in Figure~\ref{fig10}(i) and (j) we have shown the allowed values of $|M^{}_{ee}|$ as functions of the lightest neutrino mass in the parameter space for successful leptogenesis. We see that in the NO case it is below 0.006 eV and even might be vanishingly small for $m^{}_1 \lesssim 0.005$ eV, which have no chance to be probed by forseeable neutrinoless double beta decay experiments. But it can be close to 0.1 eV for $m^{}_1 \sim 0.1$ eV, which have the potential to be probed by on-going neutrinoless double beta decay experiments such as LEGEND-200 \cite{LEGEND}, KamLAND-Zen-800 \cite{KamL} and SNO+I \cite{SNO}. In the IO case it is within the range 0.02---0.05 eV for $m^{}_3 \lesssim 0.01 $ eV and can be close to 0.1 eV for $m^{}_3 \sim 0.1$ eV, which also have the potential to be probed by on-going neutrinoless double beta decay experiments.

%%%%%%%%%%%%%%%%%%%%%% FIG 1%%%%%%%%%%%%%%%%%%%%%%
%%%%%%%%%%%%%%%%%%%%%%%%%%%%%%%%%%%%%%%%%%%%%%%%%%

\subsection{Study for nearly degenerate right-handed neutrino masses}

Then, let us perform the study for the case that the right-handed neutrino masses are nearly degenerate. In the unflavored regime, the RGE induced non-zero $\varepsilon^{}_{I}$ are also suppressed by  $\Delta^2_\tau$ so that the observed value of $Y^{}_{\rm B}$ cannot be reached.

In the flavored regimes, the RGE induced non-zero $\varepsilon^{}_{1 \alpha}$ arise as
\begin{eqnarray}
&&\hspace{-1.0cm} \varepsilon^{}_{1 e} \simeq \Delta_\tau \frac{1 }{\pi v^2} \{M_2 m_2 c^2_{13} c_{12}  s_{12}  \sin (\rho -\sigma)\left[  c^2_{12}c_{23} s_{23} s_{13} \cos (\delta +\rho -\sigma)- s^2_{12}c_{23} s_{23} s_{13}\cos (\delta -\rho +\sigma) \right. \nonumber \\
&& \left. + \left(c^2_{23}s^2_{13} - s^2_{23}  \right) c_{12}  s_{12} \cos (\rho -\sigma)\right]\cdot \frac{M_1\Delta M_{21}}{4{\left( {\Delta M_{21}} \right)^2} + \Gamma^2 _2} - M_3 m_3 c^2_{13} c_{23}s_{13} c_{12}\sin (\delta +\rho) \nonumber \\
&& \left. \times \left[c_{23}s_{13} c_{12} \cos (\delta+\rho )-s_{23} s_{12}\cos \rho \right] \frac{M_1\Delta M_{31}}{4{\left( {\Delta M_{31}} \right)^2} + \Gamma^2 _3}  \right\} \;, \nonumber \\
&&\hspace{-1.0cm} \varepsilon^{}_{1 \mu} \simeq \Delta_\tau \frac{1}{2\pi v^2} \{  M_2 m_2 \left[  s^4_{12}  c^2_{23}  s^2_{23} s^2_{13} \sin 2(\delta -\rho + \sigma) -  c^4_{12}  c^2_{23}  s^2_{23} s^2_{13} \sin 2(\delta +\rho - \sigma) - c^3_{12}  c_{23}  s_{23} s_{13} s_{12} \right. \nonumber \\
&& \times \left.\left(c^2_{23}+c^2_{23}s^2_{13} - s^2_{23}  \right)\sin (\delta +2\rho - 2\sigma)  - s^3_{12}  c_{23}  s_{23} s_{13} c_{12}\left(c^2_{23} - s^2_{23}  \right)\left(1 + s^2_{13}  \right) \sin (\delta -2\rho + 2\sigma)  \right. \nonumber \\
&& - c^2_{12} s^2_{12}\left.\left( s^2_{13}- c^2_{23} s^2_{23} - s^4_{13} c^2_{23}s^2_{23}-4c^2_{23}s^2_{23}s^2_{13}  \right)\sin 2(\rho - \sigma)+c_{23}  s_{23} s_{13} c_{12}s_{12}\left(c^2_{13}s^2_{12} +c^2_{12} \right. \right. \nonumber \\
&& \left. \left. -c^2_{23} s^2_{13} c^2_{12}\right)\sin \delta +2s^3_{23} s^3_{13} c^3_{12} c_{23}s_{12}\sin (\rho - \sigma)\cos (\delta +\rho-\sigma ) \right] \cdot \frac{M_1\Delta M_{21}}{4{\left( {\Delta M_{21}} \right)^2} + \Gamma^2 _2}  \nonumber \\
&& + M_3 m_3 c^2_{13}c_{23} s_{23} \left[-s^2_{12}c_{23} s_{23}\sin 2\rho+ s^2_{13} c^2_{12}c_{23} s_{23} \sin 2(\delta +\rho )- 2s^2_{23} s_{13}c_{12} s_{12}\cos \rho \sin (\delta +\rho ) \right. \nonumber \\
&& \left. \left. +2 c^2_{23} s_{13}c_{12} s_{12}\sin \rho\cos (\delta +\rho )\right] \cdot \frac{M_1\Delta M_{31}}{4{\left( {\Delta M_{31}} \right)^2} + \Gamma^2_3} \right\} \;, \nonumber \\
&&\hspace{-1.0cm} \varepsilon^{}_{1 \tau} \simeq \Delta_\tau \frac{1}{2 \pi v^2} \{M_2 m_2 \left[  c^4_{12} c^2_{23} s^2_{23} s^2_{13} \sin 2(\delta +\rho -\sigma)-  s^4_{12} c^2_{23} s^2_{23} s^2_{13} \sin 2(\delta -\rho +\sigma)+2\left(c^2_{23}s^2_{13} - s^2_{23}  \right) \right. \nonumber \\
&& \left. \times c^3_{12} c_{23}s_{23} s_{13} s_{12}\sin(\delta+2\rho -2\sigma)+2\left(c^2_{23}s^2_{13} - s^2_{23}  \right)s^3_{12} c_{23}s_{23} s_{13} c_{12}\sin(\delta-2\rho +2\sigma) \right. \nonumber \\
&& \left. +\left(c^4_{23}s^4_{13} + s^4_{23}-4 c^2_{23}s^2_{23}s^2_{13} \right) c^2_{12}s^2_{12}\sin 2( \rho -\sigma)\right]\cdot \frac{M_1\Delta M_{21}}{4{\left( {\Delta M_{21}} \right)^2} + \Gamma^2 _2} + M_3 m_3 c^2_{23}c^2_{13} \left[s^2_{23} s^2_{12}\sin 2\rho \right. \nonumber \\
&& \left. \left. +c^2_{23} s^2_{13}c^2_{12}\sin 2(\delta+\rho)-2 c_{23} s_{23} s_{13} c_{12}  s_{12} \sin(\delta+2\rho)\right]\cdot \frac{M_1\Delta M_{31}}{4{\left( {\Delta M_{31}} \right)^2} + \Gamma^2 _3}  \right\} \;,
\end{eqnarray}
while the results of $\varepsilon^{}_{2 \alpha}$ and $\varepsilon^{}_{3 \alpha}$ are similar and collected in the appendix.
For the possible right-handed neutrino mass spectrum $M^{}_1 \approx M^{}_2 \approx M^{}_3$, the final baryon asymmetry is given by
\begin{eqnarray}
Y^{}_{\rm B}  = - c r  \left[ (\varepsilon^{}_{1\gamma} + \varepsilon^{}_{2\gamma} + \varepsilon^{}_{3\gamma})  \kappa (\widetilde m^{}_{1 \gamma} + \widetilde m^{}_{2 \gamma} + \widetilde m^{}_{3 \gamma} )  + (\varepsilon^{}_{1\tau} + \varepsilon^{}_{2\tau} + \varepsilon^{}_{3\tau})  \kappa (\widetilde m^{}_{1 \tau} + \widetilde m^{}_{2 \tau} + \widetilde m^{}_{3 \tau} )  \right] \;,
\label{6.2.4}
\end{eqnarray}
in the 2-flavor regime, and
\begin{eqnarray}
&& Y^{}_{\rm B}  = - c r \left[ (\varepsilon^{}_{1e} + \varepsilon^{}_{2e} + \varepsilon^{}_{3e})  \kappa (\widetilde m^{}_{1 e} + \widetilde m^{}_{2 e} + \widetilde m^{}_{3 e} ) + (\varepsilon^{}_{1\mu} + \varepsilon^{}_{2\mu} + \varepsilon^{}_{3\mu})  \kappa (\widetilde m^{}_{1 \mu} + \widetilde m^{}_{2 \mu} + \widetilde m^{}_{3 \mu} ) \right. \nonumber \\
&& \hspace{1.8cm} \left. + (\varepsilon^{}_{1\tau} + \varepsilon^{}_{2\tau} + \varepsilon^{}_{3\tau})  \kappa (\widetilde m^{}_{1 \tau} + \widetilde m^{}_{2 \tau} + \widetilde m^{}_{3 \tau} ) \right] \; ,
\label{6.2.5}
\end{eqnarray}
in the 3-flavor regime.
For the possible right-handed neutrino mass spectrum $M^{}_1 \ll M^{}_2 \approx M^{}_3$, the final baryon asymmetry can be calculated as in Eqs.~(\ref{5.1.2}, \ref{5.2.1}) (for the 2-flavor and 3-flavor regimes, respectively). For the possible right-handed neutrino mass spectrum $M^{}_1 \approx M^{}_2 \ll M^{}_3$, the final baryon asymmetry can be calculated as in Eqs.~(\ref{5.1.7}, \ref{5.2.4}) (for the 2-flavor and 3-flavor regimes, respectively).

%%%%%%%%%%%%%%%%%%%%%% FIG 1%%%%%%%%%%%%%%%%%%%%%%
\begin{figure*}
\centering
\includegraphics[width=6.5in]{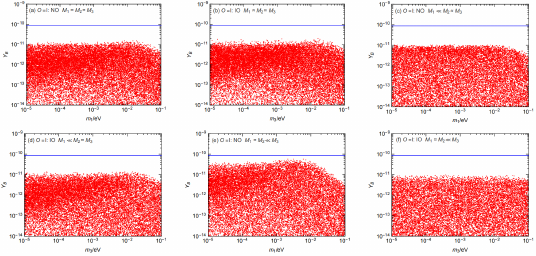}
\caption{ For the scenario studied in section~6.2, in the 2-flavor regime, the allowed values of $Y^{}_{\rm B}$ as functions of the lightest neutrino mass. }
\label{fig11}
\end{figure*}
%%%%%%%%%%%%%%%%%%%%%%%%%%%%%%%%%%%%%%%%%%%%%%%%%%

%%%%%%%%%%%%%%%%%%%%%% FIG 1%%%%%%%%%%%%%%%%%%%%%%
\begin{figure*}
\centering
\includegraphics[width=6.5in]{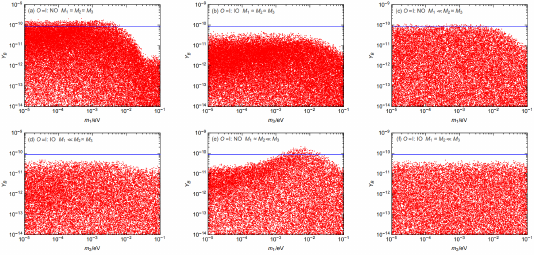}
\caption{ For the scenario studied in section~6.2, in the 3-flavor regime, the allowed values of $Y^{}_{\rm B}$ as functions of the lightest neutrino mass. }
\label{fig12}
\end{figure*}
%%%%%%%%%%%%%%%%%%%%%%%%%%%%%%%%%%%%%%%%%%%%%%%%%%

\begin{figure*}
\centering
\includegraphics[width=6.5in]{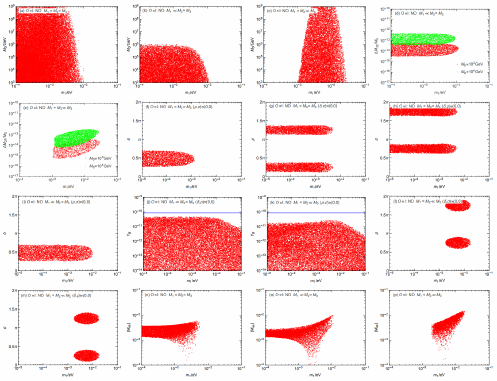}
\caption{ Some further results for the scenario studied in section~6.2 that allow for a successful leptogenesis.
(a)-(c): the values of $M^{}_0$ (as functions of $m^{}_1$) that allow for a successful leptogenesis; (d)-(e): for the benchmark values of $M^{}_0 = 10^{3}$ (red) and $10^{4}$ (green) GeV, the values of $\Delta M^{}_{32}/M^{}_0$ and $\Delta M^{}_{21}/M^{}_0$ (as functions of $m^{}_1$) that allow for a successful leptogenesis;
(f)-(i), (l)-(m): the relevant parameter space of $\delta$, $\rho$ and $\sigma$ versus $m^{}_1$ in the cases that $\delta$, $\rho$ and $\sigma$ are respectively the only source for CP violation; (j)-(k): the allowed values of $Y^{}_{\rm B}$ as functions of $m^{}_1$ in the cases that $\sigma$ and $\delta$ are respectively the only source for CP violation;
(n)-(p): the allowed values of $|M^{}_{ee}|$ as functions of $m^{}_1$ in the parameter space for successful leptogenesis. }
\label{fig13}
\end{figure*}

For the possibilities of $M^{}_1 \approx M^{}_2 \approx M^{}_3$, $M^{}_1 \ll M^{}_2 \approx M^{}_3$ and $M^{}_1 \approx M^{}_2 \ll M^{}_3$ combined with the NO and IO cases, Figure~\ref{fig11} has shown the allowed values of $Y^{}_{\rm B}$ as functions of the lightest neutrino mass in the 2-flavor regime. One can see that the maximally allowed values of $Y^{}_{\rm B}$ are smaller than the observed value by about one order of magnitude. On the other hand, Figure~\ref{fig12} has shown the results in the 3-flavor regime. Thanks to the enhanced RGE effects (due to the enlargement of the RGE energy gap), for all the possibilities of $M^{}_1 \approx M^{}_2 \approx M^{}_3$,  $M^{}_1 \ll M^{}_2 \approx M^{}_3$ and $M^{}_1 \approx M^{}_2 \ll M^{}_3$, the observed value of $Y^{}_{\rm B}$ can be reached in some parameter region in the NO case (but not in the IO case).
For these possibilities, in Figure~\ref{fig13}(a)-(c) we have shown the values of $M^{}_0$ (as functions of $m^{}_1$) that allow for a successful leptogenesis. The results show that for the possibilities of $M^{}_1 \approx M^{}_2 \approx M^{}_3$ and $M^{}_1 \approx M^{}_2 \ll M^{}_3$ a successful leptogenesis can be achieved for $M^{}_0$ in the whole temperature range of the 3-flavor regime (i.e., $\lesssim 10^{9}$ GeV), but for the possibility of $M^{}_1 \ll M^{}_2 \approx M^{}_3$ a successful leptogenesis can be achieved only for $M^{}_0 \lesssim 10^6$ GeV. And in Figure~\ref{fig13}(d) and (e) for the possibilities of $M^{}_1 \ll M^{}_2 \approx M^{}_3$ and $M^{}_1 \approx M^{}_2 \ll M^{}_3$ we have shown the values of $\Delta M^{}_{32}/M^{}_0$ and $\Delta M^{}_{21}/M^{}_0$ (as functions of $m^{}_1$) that allow for a successful leptogenesis, for the benchmark values of $M^{}_0=10^{3}$ (red) and $10^{4}$ (green) GeV. For $M^{}_0=10^{3}$ ($10^{4}$) GeV,
in order to achieve a successful leptogenesis, $\Delta M^{}_{32}/M^{}_0$ and $\Delta M^{}_{21}/M^{}_0$ should be within the range $10^{-15}$---$10^{-13}$ ($10^{-14}$---$10^{-12}$).
Then, we consider the interesting possibilities that only one of $\delta$, $\rho$ and $\sigma$ is the source for CP violation. In Figure~\ref{fig13}(f)-(h) we have shown the results for the possibility of $M^{}_1 \approx M^{}_2 \approx M^{}_3$: in the case that $\delta$ is the only source for CP violation, in order to accommodate a successful leptogenesis, $\delta$ should be around $\pi/2$ and $m^{}_1$ should be $\lesssim 0.001$ eV; in the case that $\rho$ ($\sigma$) is the only source for CP violation, in order to accommodate a successful leptogenesis, $\rho$ ($\sigma$) should be around $\pi/4$ or $5\pi/4$ ($3\pi/4$ or $7\pi/4$) and $m^{}_1$ should be $\lesssim 0.005$ eV. In Figure~\ref{fig13}(i) and (j) we have shown the results for the possibility of $M^{}_1 \ll M^{}_2 \approx M^{}_3$: in the case that $\delta$ is the only source for CP violation, in order to accommodate a successful leptogenesis, $\delta$ should be around $\pi/2$ and $m^{}_1$ should be $\lesssim 0.01$ eV; but in the case that $\sigma$ is the only source for CP violation the maximally allowed values of $Y^{}_{\rm B}$ cannot reach the observed value. In Figure~\ref{fig13}(k)-(m) we have shown the results for the possibility of $M^{}_1 \approx M^{}_2 \ll M^{}_3$: in the case that $\delta$ is the only source for CP violation the maximally allowed values of $Y^{}_{\rm B}$ cannot reach the observed value; in the case that $\rho$ ($\sigma$) is the only source for CP violation, in order to accommodate a successful leptogenesis, $\rho$ ($\sigma$) should be around $3\pi/4$ or $7\pi/4$ ($\pi/4$ or $5\pi/4$) and $m^{}_1$ should be around $0.01$ eV.
Finally, in Figure~\ref{fig13}(n)-(p) we have shown the allowed values of $|M^{}_{ee}|$ as functions of $m^{}_1$ in the parameter space for successful leptogenesis. We see that for the possibility of $M^{}_1 \approx M^{}_2 \approx M^{}_3$ it is below 0.006 eV and even might be vanishingly small for $m^{}_1 \sim 0.002$ eV, which have no chance to be probed by forseeable neutrinoless double beta decay experiments. For the possibility of $M^{}_1 \ll M^{}_2 \approx M^{}_3$ it is below 0.006 eV and even might be vanishingly small for $m^{}_1 \lesssim 0.005$ eV, which have no chance to be probed by forseeable neutrinoless double beta decay experiments, but it can be close to 0.01 eV for $m^{}_1 \sim 0.01$ eV, which have the potential to be probed by the planned of neutrinoless double beta decay experiments such as LEGEND-1000 \cite{LEGEND} and nEXO \cite{EXO}.
For the possibility of $M^{}_1 \approx M^{}_2 \ll M^{}_3$
it can exceed 0.01 eV for $m^{}_1 \sim 0.01$ eV, which have the potential to be probed by the planned of neutrinoless double beta decay experiments such as LEGEND-1000 \cite{LEGEND} and nEXO \cite{EXO}.

%%%%%%%%%%%%%%%%%%%%%% FIG 1%%%%%%%%%%%%%%%%%%%%%%
%%%%%%%%%%%%%%%%%%%%%%%%%%%%%%%%%%%%%%%%%%%%%%%%%%

\subsection{Study for scenario of $M \propto 1/m$}

As mentioned in section~3.1, in the flavor-symmetry models where the Dirac neutrino matrix is proportional to the identity matrix while the right-handed neutrino mass matrix is non-diagonal, the former will become proportional to a unitary matrix (corresponding to $O=I$) after one goes back to the basis with the latter being diagonal via a unitary transformation of the right-handed neutrinos. In this class of models, the right-handed neutrino masses are inversely proportional to the light neutrino masses, and their ratios are fixed by the corresponding light neutrino masses. For this scenario, in this subsection we study if the RGE induced leptogenesis can successfully reproduce the observed value of $Y^{}_{\rm B}$.

In the NO case (for which one has $m^{}_1 < m^{}_2 < m^{}_3$), the right-handed neutrino that is responsible for the generation of $m^{}_3$ is the lightest one, so the final baryon asymmetry mainly comes from it. In the IO case (for which one has $m^{}_3 < m^{}_1 \simeq m^{}_2$), due to the approximate equality of $m^{}_1$ and $m^{}_2$, the two right-handed neutrinos that are responsible for the generation of them are nearly degenerate, so their contributions to the final baryon asymmetry (and also the washout effects) should be taken into consideration altogether. But it should be noted that these two right-handed neutrinos are not nearly degenerate enough to give rise to a resonant leptogenesis.

\begin{figure*}
\centering
\includegraphics[width=6.5in]{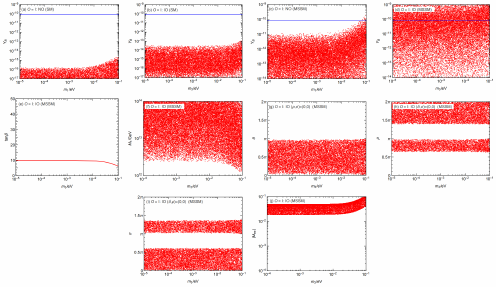}
\caption{ Some results for the scenario studied in section~6.3. (a)-(b): the allowed values of $Y^{}_{\rm B}$ as functions of the lightest neutrino mass in the SM framework; (c)-(d): the allowed values of $Y^{}_{\rm B}$ as functions of the lightest neutrino mass in the MSSM framework; (e): the minimal values of $\tan \beta$ needed to accommodate a successful leptogenesis;
(f): the values of $M^{}_1$ (as functions of $m^{}_3$) that allow for a successful leptogenesis;
(g)-(i): the relevant parameter space of $\delta$, $\rho$ and $\sigma$ versus $m^{}_3$ in the cases that $\delta$, $\rho$ and $\sigma$ are respectively the only source for CP violation;
(j): the allowed values of $|M^{}_{ee}|$ as functions of $m^{}_3$ in the parameter space for successful leptogenesis.}
\label{fig14}
\end{figure*}

Similar to the scenario studied in section~6.1, a successful RGE induced leptogenesis can only be possible in the 2-flavor regime (in consideration that the CP asymmetries are suppressed by $\Delta^2_\tau$ in the unflavored regime and by a lower leptogenesis scale in the 3-flavor regime). For the 2-flavor regime,
Figure~\ref{fig14}(a) and (b) have shown the allowed values of $Y^{}_{\rm B}$ as functions of the lightest neutrino mass in the SM framework. One can see that the maximally allowed values of $Y^{}_{\rm B}$ are smaller than the observed value by about 6 (4) orders of magnitude in the NO (IO) case. But, as shown in Figure~\ref{fig14}(c) and (d), the observed value of $Y^{}_{\rm B}$ can be reached in the IO case in the MSSM framework. For this case, Figure~\ref{fig14}(e) has shown the minimal values of $\tan \beta$ (it is about 10) needed to accommodate a successful leptogenesis.
And in Figure~\ref{fig14}(f) we have shown the values of $M^{}_1$ (as functions of $m^{}_3$) that allow for a successful leptogenesis. The results show that a successful leptogenesis can be achieved for $M^{}_1 \gtrsim 10^{12}$ GeV.
Furthermore, in Figure~\ref{fig14}(g)-(i) we have shown the relevant parameter space of $\delta$, $\rho$ and $\sigma$ versus $m^{}_3$ in the cases that $\delta$, $\rho$ and $\sigma$ are respectively the only source for CP violation.
Finally, in Figure~\ref{fig14}(j) we have shown the allowed values of $|M^{}_{ee}|$ as functions of $m^{}_1$ in the parameter space for successful leptogenesis. We see that it is within the range 0.02---0.05 eV for $m^{}_3 \lesssim 0.01 $ eV and can be close to 0.1 eV for $m^{}_3 \sim 0.1$ eV,
which have the potential to be probed by on-going neutrinoless double beta decay experiments such as LEGEND-200 \cite{LEGEND}, KamLAND-Zen-800 \cite{KamL} and SNO+I \cite{SNO}.

%%%%%%%%%%%%%%%%%%%%%% FIG 1%%%%%%%%%%%%%%%%%%%%%%
%%%%%%%%%%%%%%%%%%%%%%%%%%%%%%%%%%%%%%%%%%%%%%%%%%

\section{Summary}

In the literature, motivated by the observed peculiar neutrino mixing pattern and a preliminary experimental hint for maximal Dirac CP phase (i.e., $\delta \sim 3\pi/2$), a lot of flavor and CP symmetries have been proposed to help us understand and explain these experimental results. However, for some flavor-symmetry scenarios, the leptogenesis mechanism (which accompanies the seesaw model and provides an elegant explanation for the baryon-antibaryon asymmetry of the Universe) is prohibited to work as usual (see section~3.1 and Table~\ref{tab2}). To tackle this problem, in this paper we have made an exhausitive study on the possibility that the renormalization group evolution effect may induce a successful leptogenesis for these particular scenarios. The motivation for such a study is twofold: this effect is spontaneous, provided that there is a considerable gap between the flavor-symmetry scale and the leptogenesis scale; this effect is minimal, in the sense that it does not need to introduce additional flavor-symmetry-breaking parameters. Our study provides some complementarities to the previous related studies in Refs.~\cite{rgeL1, rgeL2, rgeL3} (see section~3.2).

The flavor-symmetry scenarios we have studied and the main results are summarized as follows:

In section~4, for the scenarios of $O =O^{}_1$, $O^{}_2$, $O^{}_3$ and $O^{}_4$ [see Eqs.~(\ref{3.1.2}, \ref{3.1.3}) for their definitions] and the unflavored regime, we study if the RGE induced leptogenesis can successfully reproduce the observed value of $Y^{}_{\rm B}$ in both the cases that the right-handed neutrino masses are hierarchical and nearly degenerate. For simplicity and clarity, we have just considered the cases that only one of the parameters $x$, $y$ and $z$ in Eq.~(\ref{3.1.2}) is non-vanishing (i.e., the scenarios of $O=O^{}_x, O^{}_y, O^{}_z, O^{\prime}_x, O^{\prime}_y$ and $O^{\prime}_z$). In the case that the right-handed neutrino masses are hierarchical, the RGE induced values of $Y^{}_{\rm B}$ cannot reach the observed value in the SM framework. But in the MSSM framework, thanks to the enhancement of the RGE effects from a proper value of $\tan \beta$, the observed value of $Y^{}_{\rm B}$ can be reached for the scenarios of $O= O^{}_y$ (in both the NO and IO cases) and $O^{\prime}_y$ (in the NO case). In the case that the right-handed neutrino masses are nearly degenerate, for the scenarios of $O= O^{}_z$ and $O^{\prime}_z$ and the right-handed neutrino mass spectrum $M^{}_1 \approx M^{}_2 \ll M^{}_3$, the observed value of $Y^{}_{\rm B}$ can be reached in some parameter region in the NO case (but not in the IO case).

In section~5, for the scenarios of $O=O^{\prime}_x, O^{\prime}_y$ and $O^{\prime}_z$ and the flavored regimes, we study if the RGE induced leptogenesis can successfully reproduce the observed value of $Y^{}_{\rm B}$ in the case that the right-handed neutrino masses are nearly degenerate. In the 2-flavor regime, only for the scenario of $O=O^{\prime}_z$ and the right-handed neutrino mass spectrum $M^{}_1 \approx M^{}_2 \ll M^{}_3$ can the observed value of $Y^{}_{\rm B}$ be reached in some parameter region in the NO case (but not in the IO case). In the 3-flavor regime, thanks to the enhancement of the RGE effects from the enlargement of the RGE energy gap, for the scenario of $O=O^{\prime}_x$ in combination with the right-handed neutrino mass spectrum $M^{}_1 \ll M^{}_2 \approx M^{}_3$ and $M^{}_1 \approx M^{}_2 \approx M^{}_3$, and for the scenario of $O= O^{\prime}_z$ in combination with the right-handed neutrino mass spectrum $M^{}_1 \approx M^{}_2 \ll M^{}_3$ and $M^{}_1 \approx M^{}_2 \approx M^{}_3$, the observed value of $Y^{}_{\rm B}$ can be reached in some parameter region in the NO case (but not in the IO case).

In section~6, for the scenario of $O=I$ and all the three flavored regimes,  we study if the RGE induced leptogenesis can successfully reproduce the observed value of $Y^{}_{\rm B}$ in both the cases that the right-handed neutrino masses are hierarchical and nearly degenerate. In the unflavored regime, the RGE induced non-zero $\varepsilon^{}_{I}$ are suppressed by $\Delta^2_\tau$ so that the observed value of $Y^{}_{\rm B}$ cannot be reached. As for the flavored regimes, in the case that the right-handed neutrino masses are hierarchical, the observed value of $Y^{}_{\rm B}$ cannot neither be reached in the SM framework. But in the MSSM framework, for proper values of $\tan \beta$, the observed value of $Y^{}_{\rm B}$ can be reached in the 2-flavor regime (but not in the 3-flavor regime).
In the case that the right-handed neutrino masses are nearly degenerate, the RGE induced values of $Y^{}_{\rm B}$ cannot reach the observed value in the 2-flavor regime. But in the 3-flavor regime, for the right-handed neutrino mass spectrum $M^{}_1 \approx M^{}_2 \approx M^{}_3$, $M^{}_1 \ll M^{}_2 \approx M^{}_3$ and $M^{}_1 \approx M^{}_2 \ll M^{}_3$, the observed value of $Y^{}_{\rm B}$ can be reached in some parameter region in the NO case (but not in the IO case). We have also considered the case that the right-handed neutrino masses are inversely proportional to the light neutrino masses. In this case, the observed value of $Y^{}_{\rm B}$ cannot be reached in the SM framework. But in the MSSM framework, for proper values of $\tan \beta$, the observed value of $Y^{}_{\rm B}$ can be reached in the 2-flavor regime and the IO case.

For all the scenarios that can accommodate a successful RGE induced leptogenesis, we have shown the relevant parameter space of the CP phases and the allowed values of the effective neutrino mass that controls the rates of neutrinoless double beta decays. And we have shown the values of the right-handed neutrino masses that allow for a successful leptogenesis. For the cases that the right-handed neutrino masses are nearly degenerate, we have also shown the values of the right-handed neutrino mass differences that allow for a successful leptogenesis.

\vspace{0.5cm}

\underline{Acknowledgments} \vspace{0.2cm}

This work is supported in part by the National Natural Science Foundation of China under grant NO.~12475112, the Natural Science Foundation of the Liaoning Scientific Committee under grant NO.~2022-MS-314, and the Basic Research Business Fees for Universities in Liaoning Province (2024).

\section*{Appendix}

In the case that the right-handed neutrino masses are nearly degenerate, the RGE induced non-zero $\varepsilon^{}_{2 \alpha}$ and $\varepsilon^{}_{3 \alpha}$ arise as
\begin{eqnarray}
&&\hspace{-1.0cm} \varepsilon^{}_{2 e} \simeq \Delta_\tau \frac{1 }{\pi v^2} \{M_1 m_1 c^2_{13} c_{12}  s_{12}  \sin (\rho -\sigma)\left[  c^2_{12}c_{23} s_{23} s_{13} \cos (\delta +\rho -\sigma)- s^2_{12}c_{23} s_{23} s_{13}\cos (\delta -\rho +\sigma) \right. \nonumber \\
&& \left. + \left(c^2_{23}s^2_{13} - s^2_{23}  \right) c_{12}  s_{12} \cos (\rho -\sigma)\right]\cdot \frac{M_1\Delta M_{21}}{4{\left( {\Delta M_{21}} \right)^2} + \Gamma^2_1} - M_3 m_3 c^2_{13} c_{23}s_{13} c_{12}\sin (\delta +\sigma) \nonumber \\
&& \left. \times \left[c_{23}s_{13} c_{12} \cos (\delta+\sigma )+s_{23} s_{12}\cos \sigma \right] \frac{M_1\Delta M_{32}}{4{\left( {\Delta M_{32}} \right)^2} + \Gamma^2 _3}  \right\} \;, \nonumber \\
&&\hspace{-1.0cm} \varepsilon^{}_{2 \mu} \simeq \Delta_\tau \frac{1}{2\pi v^2} \{  M_1 m_1 \left[  s^4_{12}  c^2_{23}  s^2_{23} s^2_{13} \sin 2(\delta -\rho + \sigma) -  c^4_{12}  c^2_{23}  s^2_{23} s^2_{13} \sin 2(\delta +\rho - \sigma) - c^3_{12}  c_{23}  s_{23} s_{13} s_{12} \right. \nonumber \\
&& \times \left.\left(c^2_{23}+c^2_{23}s^2_{13} - s^2_{23}  \right)\sin (\delta +2\rho - 2\sigma)  - s^3_{12}  c_{23}  s_{23} s_{13} c_{12}\left(c^2_{23} - s^2_{23}  \right)\left(1 + s^2_{13}  \right) \sin (\delta -2\rho + 2\sigma)  \right. \nonumber \\
&& - c^2_{12} s^2_{12}\left.\left( s^2_{13}- c^2_{23} s^2_{23} - s^4_{13} c^2_{23}s^2_{23}-4c^2_{23}s^2_{23}s^2_{13}  \right)\sin 2(\rho - \sigma)+c_{23}  s_{23} s_{13} c_{12}s_{12}\left(c^2_{13}s^2_{12} +c^2_{12} \right. \right. \nonumber \\
&& \left. \left. -c^2_{23} s^2_{13} c^2_{12}\right)\sin \delta +2s^3_{23} s^3_{13} c^3_{12} c_{23}s_{12}\sin (\rho - \sigma)\cos (\delta +\rho-\sigma ) \right] \cdot \frac{M_1\Delta M_{21}}{4{\left( {\Delta M_{21}} \right)^2} + \Gamma^2_1}  \nonumber \\
&& + M_3 m_3 c^2_{13}c_{23} s_{23} \left[-c^2_{12}c_{23} s_{23}\sin 2\sigma+ s^2_{13} s^2_{12}c_{23} s_{23} \sin 2(\delta +\sigma ) + 2s^2_{23} s_{13}c_{12} s_{12}\cos \sigma \sin (\delta +\sigma) \right. \nonumber \\
&& \left. \left. -2 c^2_{23} s_{13}c_{12} s_{12}\sin \sigma \cos (\delta +\sigma )\right] \cdot \frac{M_1\Delta M_{32}}{4{\left( {\Delta M_{32}} \right)^2} + \Gamma^2_3} \right\} \;, \nonumber \\
&&\hspace{-1.0cm} \varepsilon^{}_{2 \tau} \simeq \Delta_\tau \frac{1}{2 \pi v^2} \{M_1 m_1 \left[  c^4_{12} c^2_{23} s^2_{23} s^2_{13} \sin 2(\delta +\rho -\sigma)-  s^4_{12} c^2_{23} s^2_{23} s^2_{13} \sin 2(\delta -\rho +\sigma)+2\left(c^2_{23}s^2_{13} - s^2_{23}  \right) \right. \nonumber \\
&& \left. \times c^3_{12} c_{23}s_{23} s_{13} s_{12}\sin(\delta+2\rho -2\sigma)+2\left(c^2_{23}s^2_{13} - s^2_{23}  \right)s^3_{12} c_{23}s_{23} s_{13} c_{12}\sin(\delta-2\rho +2\sigma) \right. \nonumber \\
&& \left. +\left(c^4_{23}s^4_{13} + s^4_{23}-4 c^2_{23}s^2_{23}s^2_{13} \right) c^2_{12}s^2_{12}\sin 2( \rho -\sigma)\right]\cdot \frac{M_1\Delta M_{21}}{4{\left( {\Delta M_{21}} \right)^2} + \Gamma^2 _1} + M_3 m_3 c^2_{23}c^2_{13} \left[s^2_{23} c^2_{12}\sin 2\sigma \right. \nonumber \\
&& \left. \left. +c^2_{23} s^2_{13} s^2_{12} \sin 2(\delta+\sigma) + 2 c_{23} s_{23} s_{13} c_{12}  s_{12} \sin(\delta+2\sigma)\right]\cdot \frac{M_1\Delta M_{32}}{4{\left( {\Delta M_{32}} \right)^2} + \Gamma^2 _3}  \right\} \;,
\end{eqnarray}
\begin{eqnarray}
&&\hspace{-1.0cm} \varepsilon^{}_{3 e} \simeq -\Delta_\tau \frac{1}{\pi v^2} \{M_1 m_1 c^2_{13} c_{23} s_{13} c_{12} \sin (\delta +\rho) \left[ c_{23} s_{13} c_{12} \cos (\delta +\rho )-  s_{23} s_{12}\cos \rho \right]\cdot \frac{M_1\Delta M_{31}}{4{\left( {\Delta M_{31}} \right)^2} + \Gamma^2 _1}  \nonumber \\
&& + M_2 m_2 c^2_{13}c_{23}s_{13} s_{12} \sin (\delta +\sigma) \left[c_{23}s_{13} s_{12} \cos (\delta+\sigma )+s_{23} c_{12}\cos \sigma \right]\cdot \frac{M_1\Delta M_{32}}{4{\left( {\Delta M_{32}} \right)^2} + \Gamma^2 _2}  \}\;, \nonumber \\
&&\hspace{-1.0cm} \varepsilon^{}_{3 \mu} \simeq -\Delta_\tau \frac{1 }{2 \pi v^2} \{M_1 m_1 c^2_{13} c_{23} s_{23}\left[s^2_{12} c_{23} s_{23}  \sin 2\rho- s^2_{13} c^2_{12} c_{23} s_{23} \sin 2(\delta +\rho)  +2 s^2_{23}s_{13} c_{12}s_{12} \cos \rho  \right. \nonumber \\
&& \left. \times \sin (\delta +\rho )-2c^2_{23}s_{13} c_{12}s_{12} \sin \rho  \cos (\delta +\rho )\right] \cdot \frac{M_1\Delta M_{31}}{4{\left( {\Delta M_{31}} \right)^2} + \Gamma^2 _1}  + M_2 m_2 c^2_{13}  c_{23} s_{23}\left[ c^2_{12} c_{23} s_{23}  \sin 2\sigma  \right. \nonumber \\
&& \left. - 2 s^2_{23}s_{13} c_{12}s_{12} \cos \sigma  \sin (\delta +\sigma )+2c^2_{23}s_{13} c_{12}s_{12} \sin \sigma\cos (\delta +\sigma ) -s^2_{13}s^2_{12}c_{23} s_{23}\sin 2(\delta +\sigma )\right] \nonumber \\
&& \cdot \left. \frac{M_1\Delta M_{32}}{4{\left( {\Delta M_{32}} \right)^2} + \Gamma^2 _2}  \right\} \nonumber \\
&&\hspace{-1.0cm} \varepsilon^{}_{3 \tau} \simeq \Delta_\tau \frac{1 }{2 \pi v^2} \{M_1 m_1 c^2_{23}c^2_{13}\left[ c^2_{23}  s^2_{13}c^2_{12} \sin 2(\delta +\rho )+  s^2_{23} s^2_{12} \sin 2\rho -2c_{23} s_{23} s_{13} c_{12}s_{12}\sin (\delta +2\rho )\right] \nonumber \\
&& \cdot \frac{M_1\Delta M_{31}}{4{\left( {\Delta M_{31}} \right)^2} + \Gamma^2 _1}+ M_2 m_2 c^2_{23}c^2_{13} \left[s^2_{23} c^2_{12}\sin 2\sigma+c^2_{23} s^2_{13}s^2_{12}\sin 2(\delta+\sigma)+2 c_{23} s_{23} s_{13} c_{12}s_{12} \right. \nonumber \\
&& \left. \sin(\delta+2\sigma)\right]\cdot \frac{M_1\Delta M_{32}}{4{\left( {\Delta M_{32}} \right)^2} + \Gamma^2 _2}  \} \;.
\label{a2}
\end{eqnarray}

\end{document}